\documentclass[twocolumn]{aastex62}

\received{}
\revised{}
\accepted{}

\submitjournal{The Astrophysical Journal}

\usepackage{rotating}
\usepackage{gensymb}

\shorttitle{Electron Preacceleration in Weak ICM Shocks}
\shortauthors{Kang et al.}

\begin{document}

\title{Electron Preacceleration in Weak Quasi-perpendicular Shocks in High-beta Intracluster Medium}

\author[0000-0002-4674-5687]{Hyesung Kang}
\affiliation{Department of Earth Sciences, Pusan National University, Busan 46241, Korea}
\author[0000-0002-5455-2957]{Dongsu Ryu}
\affiliation{Department of Physics, School of Natural Sciences UNIST, Ulsan 44919, Korea}
\author[0000-0001-7670-4897]{Ji-Hoon Ha}
\affil{Department of Physics, School of Natural Sciences UNIST, Ulsan 44919, Korea}
\correspondingauthor{Hyesung Kang}
\email{hskang@pusan.ac.kr}

\begin{abstract}


Giant radio relics in the outskirts of galaxy clusters are known to be lit up by the relativistic electrons produced via diffusive shock acceleration (DSA) in shocks with low sonic Mach numbers, $M_{\rm s}\lesssim3$. 
The particle acceleration at these collisionless shocks critically depends on the kinetic plasma processes that govern the injection to DSA.
Here, we study the preacceleration of suprathermal electrons in weak, quasi-perpendicular ($Q_\perp$) shocks in the hot, high-$\beta$ ($\beta = P_{\rm gas}/P_{\rm B}$) intracluster medium (ICM) through two-dimensional particle-in-cell simulations. \citet{guo2014a,guo2014b} showed that in high-$\beta$ $Q_\perp$-shocks, some of incoming electrons could be reflected upstream and gain energy via shock drift acceleration (SDA). The temperature anisotropy due to the SDA-energized electrons then induces the electron firehose instability (EFI), and oblique waves are generated, leading to a Fermi-like process and multiple cycles of SDA in the preshock region. 
We find that such electron preacceleration is effective only in shocks above a critical Mach number $M_{\rm ef}^*\approx2.3$.
This means that in ICM plasmas, $Q_\perp$-shocks with $M_{\rm s}\lesssim2.3$ may not efficiently accelerate electrons.
We also find that even in $Q_\perp$-shocks with $M_{\rm s}\gtrsim2.3$, electrons may not reach high enough energies to be injected to the full Fermi-I process of DSA, because long-wavelength waves are not developed via the EFI alone.
{\color{black} Our results indicate that additional electron preaccelerations are required for DSA in ICM shocks, and the presence of fossil relativistic electrons in the shock upstream region may be necessary to explain observed radio relics.}

\end{abstract}


\section{Introduction} 
\label{sec:s1}

Weak shocks with low sonic Mach numbers, $M_{\rm s}\lesssim 3$, form in the hot intracluster medium (ICM) during major merges of galaxy clusters \citep[e.g.,][]{gabici2003,ryu2003, ha2018a}.
Radiative signatures of those merger shocks have been detected in X-ray and radio observations \citep[e.g.,][]{markevitch07,vanweeren10,bruggen2012,brunetti2014}.
In the case of the so-called radio relics, the radio emission has been interpreted as the synchrotron radiation from the relativistic electrons accelerated via diffusive shock acceleration (DSA) in the shocks. 
Hence, the sonic Mach numbers of relic shocks, $M_{\rm radio}$ (radio Mach number), have been inferred from the radio spectral index \citep[e.g.,][]{vanweeren10,vanweeren2016}, based on the DSA test-particle power-law energy spectrum \citep[e.g.,][]{bell1978,blandford1978,drury1983}.
In X-ray observations, the sonic Mach numbers, $M_{\rm X}$ (X-ray Mach number), have been estimated for merger-driven shocks, using the discontinuities in temperature or surface brightness \citep[e.g.,][]{markevitch2002,markevitch07}.

While $M_{\rm radio}$ and $M_{\rm X}$ are expected to match, $M_{\rm radio}$ has been estimated to be larger than $M_{\rm X}$ in some radio relics \citep[e.g.,][]{akamatsu2013}. 
In the case of the Toothbrush radio relic in merging cluster 1RXS J060303.3, for instance, \citet{vanweeren2016} estimated that $M_{\rm radio}\approx 2.8$ while $M_{\rm X}\approx 1.2-1.5$.
In the so-called reaccelertion model, weak shocks with $\sim M_{\rm X}$ are presumed to sweep through fossil electrons with power-law energy spectrum, $N_{\rm fossil} \propto \gamma^{-p}$ ($\gamma$ is the Lorentz factor), and then the radio spectra with observed spectral indices, $\alpha_{\rm sh} = (p-1)/2$, are supposed to be generated \citep[e.g.,][]{kang16a,kang16b}.
This model may explain the discrepancy between $M_{\rm radio}$ and $M_{\rm X}$ in some cases.
However, it may not be realistic to assume the presence of fossil electrons with flat power-law spectra up to $\gamma\sim 10^4$ over length scales of $400-500$~kpc, since such high-energy electrons cool with time scales of $\sim 100$~Myr \citep{kangetal17}. On the other hand, with mock X-ray and radio observations of radio relics using simulated clusters, \citet{hong15} argued that the surfaces of merger shocks are highly inhomogeneous in terms of $M_{\rm s}$ \citep[see, also][]{ha2018a}, and X-ray observations preferentially pick up the parts with lower $M_{\rm s}$ (higher shock energy flux), while radio emissions manifest the parts with higher $M_{\rm s}$ (higher electron acceleration). As a result, $M_{\rm X}$ could be be smaller than $M_{\rm radio}$. However, the true origins of this discrepancy have yet to be understood.

For the full description of radio relics, hence, it is necessary to first understand shocks in the ICM.
They are collisionless shocks, as in other astrophysical environments \citep[e.g.,][]{brunetti2014}.
The physics of collisionless shocks involves complex kinetic plasma processes well beyond the MHD Rankine-Hugoniot jump condition.
DSA, for instance, depends on various shock parameters including the sonic Mach number, $M_{\rm s}$, the plasma beta, $\beta = P_{\rm gas}/P_{\rm B}$ (the ratio of thermal to magnetic pressures), and the obliquity angle between the upstream background magnetic field direction and the shock normal, $\theta_{\rm Bn}$ \citep[see][]{balogh13}.

In general, collisionless shocks can be classified by the obliquity angle as {\it quasi-parallel} ($Q_\parallel$, hereafter) shocks with $\theta_{\rm Bn}\lesssim 45^{\circ}$ and {\it quasi-perpendicular} ($Q_\perp$, hereafter) shocks with $\theta_{\rm Bn}\gtrsim 45^{\circ}$.
In situ observations of Earth's bow shock indicate that protons are effectively accelerated at the $Q_\parallel$-portion, while electrons are energized preferentially in the $Q_\perp$-configuration \citep[e.g.,][]{gosling1989}.
In such shocks, one of key processes for DSA is particle injection, which involves the reflection of particles at the shock ramp, the excitation of electromagnetic waves/turbulences by the reflected particles, and the energization of particles through ensuing wave-particle interactions \citep[e.g.,][]{treumann2008,treumann2009}.
Since the thickness of the shock transition zone is of the order of the gyroradius of postshock thermal ions,
both ions and electrons need to be preaccelerated to suprathermal momenta greater than a few times the momentum of thermal ions, $p_{\rm th,i}$, in order to diffuse across the shock transition layer and fully participate in the first-order Fermi (Fermi-I, hereafter) process of DSA \citep[e.g.,][]{kang2002,caprioli2015}.
Here, $p_{\rm th,i}=\sqrt{2m_i k_B T_{i2}}$, $T_{i2}$ is the postshock ion temperature and $k_B$ is the Boltzmann constant.
Hereafter, the subscripts 1 and 2 denote the preshock and postshock quantities, respectively.

Kinetic processes in collisionless shocks can be studied through, for instance, particle-in-cell (PIC) and hybrid plasma simulations \citep[e.g.,][]{caprioli2014a,guo2014a,guo2014b,park2015}.
Previous studies have mostly focused on shocks in $\beta \lesssim 1$ plasmas, where the Aflv\'en Mach number $M_{\rm A}$ is about the same as $M_{\rm s}$ ($M_{\rm A}\approx \sqrt{\beta} M_{\rm s}$), investigating shocks in solar wind and the interstellar medium (ISM)  \citep[see also][and references therein]{treumann2009}.
If plasmas have very low-$\beta$ (sometimes referred as cold plasmas), even the thermal motions of particles can be neglected .
In hot ICM plasmas, on the other hand, $\beta \sim 100$ \citep[e.g.,][]{ryu2008,porter2015}, and shocks have low sonic Mach numbers of $M_{\rm s} \lesssim 3$, 
but relatively high Alfv\'en Mach numbers up to $M_{\rm A} \approx 30$.
In such shocks, kinetic processes are expected to operate differently from low-$\beta$ shocks.

Recently, we investigated proton acceleration in weak ($M_{\rm s} \approx 2 - 4$) ``$Q_\parallel$-shocks'' in high-$\beta$ ($\beta= 30-100$) ICM plasmas through one-dimensional (1D) and two-dimensional (2D) PIC simulations \citep[][Paper I, hereafter]{ha2018b}.
The main findings can be recapitulated as follows.
(1) $Q_\parallel$-shocks with $M_{\rm s} \gtrsim 2.3$ develop overshoot-undershoot oscillations in their structures and undergo quasi-cyclic reformation, leading to a significant amount of incoming protons being reflected at the shock. The backstreaming ions excite resonant and non-resonant waves in the foreshock region, leading to the generation of suprathermal protons that can be injected to the Fermi-I process.
(2) $Q_\parallel$-shocks with $M_{\rm s} \lesssim 2.3$, on the other hand, have relatively smooth and steady structures. The development of suprathermal population is negligible in these shocks.
(3) In $Q_\perp$-shocks, a substantial fraction of incoming ions are reflected and gain energy via shock drift acceleration (SDA), but the energized ions advect downstream along with the background magnetic field after about one gyromotion without being injected to the Fermi-I acceleration.
(4) For the description of shock dynamics and particle acceleration in high-$\beta$ plasmas, the sonic Mach number is the more relevant parameter than the Alfv\'en Mach number,  since the reflection of particles is mostly controlled by $M_{\rm s}$.

As a sequal, in this work, we explore the electron preacceleration in low Mach number, $Q_\perp$-shocks in high-$\beta$ ICM plasmas. Such shocks are thought be the agents of radio relics in merging clusters.
Previously, the pre-energization of thermal electrons at collisionless shocks (i.e., the injection problem), which involves kinetic processes such as the excitation of waves via micro-instabilities and wave-particle interactions, was studied through PIC simulations \citep[e.g.,][and references therein]{amano09,riqu11,matsukiyo11,guo2014a, guo2014b,park2015}.
For instance, \citet{amano09} showed that in high-$M_{\rm A}$ $Q_\perp$-shocks with $\beta\sim 1$, strong electrostatic waves are excited by Buneman instability and confine electrons in the shock foot region, where electrons gain energy by drifting along the motional electric field (shock surfing acceleration, SSA). 
On the other hand, \citet{riqu11} found that in $Q_\perp$-shocks with $M_{\rm A} \lesssim (m_i/m_e)^{1/2}$ (where $m_i/m_e$ is the ion-to-electron mass ratio) and $\beta\sim 1$, the growth of oblique whistler waves in the shock foot by modified two-stream instabilities (MTSIs) may play important roles in confining and pre-energizing electrons.
{\color{black} \citet{matsukiyo11} showed through 1D PIC simulations that in weak shocks with the fast Mach number, $M_f \approx 2-3$, and $\beta \approx 3$,
a fraction of incoming electrons are accelerated and reflected through SDA and form a suprathermal population.
However, this finding was refuted later by \citet{matsukiyo15} who showed through 2D PIC simulations of shocks with similar parameters that 
electron reflection is suppressed due to shock surface ripping.}

\citet[][GSN14a and GSN14b, hereafter]{guo2014a, guo2014b} carried out comprehensive studies for electron preacceleration and injection in $M_{\rm s}=3$, $Q_\perp$-shocks in plasmas with $\beta = 6-200$ using 2D PIC simulations.
In particular, GSN14a presented the ``relativistic SDA theory'' for oblique shocks, which can be briefly summarized as follows.
The incoming electrons that satisfy criteria (i.e., with pitch angles larger than the loss-cone angle) are reflected and gain energy through SDA at the shock ramp.
The energized electrons backstream along the background magnetic field lines with small pitch angles, generating the temperature anisotropy of $T_{e\parallel}> T_{e\perp}$.
GSN14b then showed that the ``electron firehose instability'' (EFI) is induced by the temperature anisotropy, and oblique waves are excited \citep{gary2003}. 
The electrons are scattered back and forth between magnetic mirrors at the shock ramp and self-generated upstream waves (a Fermi-I type process), being further accelerated mostly through SDA.
At this stage, the electrons are still suprathermal and do not have sufficient energies to diffuse downstream of the shock; 
instead, they stay upstream of the shock ramp.
The authors named this process as a ``Fermi-like process'', as opposed to the full, {\it bona fide} Fermi-I process.
GSN14a also pointed out that SSA does not operate in weak ICM shocks because of the suppression of Buneman instability in hot plasmas, and that in high-$\beta$ shocks the preacceleration via SDA dominates over the energization through interactions with the oblique whistler waves generated via MTSIs in the shock foot.

For electron preacceleration in weak ICM shocks, however, there are still issues to be further addressed.
Most of all, there should be a critical Mach number, below which the preacceleration is not efficient.
Even though electrons are pre-energized at shocks with $M_{\rm s} \approx 3$, as shown in GSN14a and GSN14b, it is not clear whether they could be further accelerated by the full Fermi-I process of DSA.
We will investigate these issues using 2D PIC simulations in this paper.

The paper is organized as follows. Section \ref{sec:s2} includes the descriptions of simulations, along with the definitions of various parameters involved.
In Section \ref{sec:s3}, we give a brief review on the background physics of $Q_\perp$-shocks, in order to facilitate the understandings of our simulation results in the following section.
Next, in Section \ref{sec:s4}, we present shock structures and electron preacceleration in simulations, and examine the dependence of our findings on various shock parameters. 
A brief summary is given in Section \ref{sec:s5}. 

\section{Numerics}
\label{sec:s2}

\begin{deluxetable*}{ccccccccccccc}[t]
\tablecaption{Model Parameters of Simulations \label{tab:t1}}
\tabletypesize{\scriptsize}
\tablecolumns{13}
\tablenum{1}
\tablewidth{0pt}
\tablehead{
\colhead{Model Name} &
\colhead{$M_{\rm s}$} &
\colhead{$M_{\rm A}$} &
\colhead{$u_0/c$} &
\colhead{$\theta_{\rm Bn}$} &
\colhead{$\beta$} &
\colhead{$T_{e1} = T_{i1} [\rm K(keV)]$} &
\colhead{$m_i/m_e$} &
\colhead{$L_x [c/w_{\rm pe}]$} &
\colhead{$L_y [c/w_{\rm pe}]$} &
\colhead{$\Delta x[c/w_{\rm pe}]$} &
\colhead{$t_{\rm end} [w_{\rm pe}^{-1}]$}&
\colhead{$t_{\rm end} [\Omega_{\rm ci}^{-1}]$}
}
\startdata
M2.0& 2.0 & 18.2 & 0.027 & $63^{\circ}$ &100 & $10^8(8.6)$ & 100 & $7\times 10^3$ & 80 & 0.1 & $1.13\times 10^5$ & 30\\
M2.15& 2.15 & 19.6 & 0.0297 & $63^{\circ}$ &100 & $10^8(8.6)$ & 100 & $7\times 10^3$ & 80 & 0.1 & $1.13\times 10^5$ & 30\\
M2.3 & 2.3 & 21 & 0.0325 & $63^{\circ}$ &100 & $10^8(8.6)$ & 100 & $7\times 10^3$ & 80 & 0.1 & $1.13\times 10^5$ & 30\\
M2.5& 2.5 & 22.9 & 0.035 & $63^{\circ}$ &100 & $10^8(8.6)$ & 100 & $7\times 10^3$ &  80 & 0.1 & $1.13\times 10^5$ & 30\\
M2.75& 2.75 & 25.1 & 0.041 & $63^{\circ}$ &100 & $10^8(8.6)$ & 100 & $7\times 10^3$ &  80 & 0.1 & $1.13\times 10^5$ & 30\\
M3.0& 3.0 & 27.4 & 0.047 & $63^{\circ}$ &100 & $10^8(8.6)$ & 100 & $1.2\times 10^4$ &  80 & 0.1 & $2.26\times 10^5$ & 60\\
\hline
M2.15-$\theta$53& 2.15 & 19.6 & 0.0297 & $53^{\circ}$ &100 & $10^8(8.6)$ & 100 & $7\times 10^3$ & 80 & 0.1 & $1.13\times 10^5$ & 30\\
M2.15-$\theta$73& 2.15 & 19.6 & 0.0297 & $73^{\circ}$ &100 & $10^8(8.6)$ & 100 & $7\times 10^3$ & 80 & 0.1 & $1.13\times 10^5$ & 30\\
M2.3-$\theta$53 & 2.3 & 21 & 0.0325  & $53^{\circ}$ &100 & $10^8(8.6)$ & 100 & $7\times 10^3$ & 80 & 0.1 & $1.13\times 10^5$ & 30\\
M2.3-$\theta$73 & 2.3 & 21 & 0.0325  & $73^{\circ}$ &100 & $10^8(8.6)$ & 100 & $7\times 10^3$ & 80 & 0.1 & $1.13\times 10^5$ & 30\\
\hline
M2.0-$\beta$50 &  2.0 & 12.9 & 0.027 & $63^{\circ}$ & 50 & $10^8(8.6)$ & 100 & $7\times 10^3$ & 80 & 0.1 & $8.0\times 10^4$ & 30\\
M2.3-$\beta$50 &  2.3 & 14.8 & 0.0325 & $63^{\circ}$ & 50 & $10^8(8.6)$ & 100 & $7\times 10^3$ & 80 & 0.1 & $8.0\times 10^4$ & 30\\
M3.0-$\beta$50 &  3.0 & 19.4 & 0.047 & $63^{\circ}$ & 50 & $10^8(8.6)$ & 100 & $7\times 10^3$ & 80 & 0.1 & $8.0\times 10^4$ & 30\\
\hline
M2.0-m400 &  2.0 & 18.2 & 0.013 & $63^{\circ}$ &100 & $10^8(8.6)$ & 400 & $7\times 10^3$ & 80 & 0.1 & $1.5\times 10^5$ & 10\\
M2.3-m400 &  2.3 & 21 & 0.016 & $63^{\circ}$ &100 & $10^8(8.6)$ & 400 & $7\times 10^3$ & 80 & 0.1 & $1.5\times 10^5$ & 10\\
M3.0-m400 &  3.0 & 27.4 & 0.023 & $63^{\circ}$ &100 & $10^8(8.6)$ & 400 & $7\times 10^3$ & 80 & 0.1 & $1.5\times 10^5$ & 10\\
\hline
M2.3-r2 & 2.3 & 21& 0.0325 & $63^{\circ}$ &100 & $10^8(8.6)$ & 100 & $7\times 10^3$ & 80 & 0.05 & $3.8\times 10^4$ & 10\\
M2.3-r0.5 &2.3 & 21 & 0.0325 & $63^{\circ}$ &100 & $10^8(8.6)$ & 100 & $7\times 10^3$ & 80 & 0.2 & $3.8\times 10^4$ & 10\\
\enddata
\end{deluxetable*}

Simulations were performed using TRISTAN-MP, a parallelized electromagnetic PIC code \citep{buneman1993, spitkovsky2005}. The geometry is 2D planar, while all the three components of particle velocity and electromagnetic fields are followed.
The details of simulation setups can be found in Paper I, and below some basic definitions of parameters and features are described in order to make this paper self-contained.
In Paper I, we used the variable $\mathbf{v}$, for example, $v_0$ and $v_{\rm sh}$, to represents flow velocities.
We here, however, use the variable $\mathbf{u}$ for ``flow'' velocities, while $\mathbf{v}$ is reserved for ``particle'' velocities.

Plasmas, which are composed of ions and electrons of Maxwellian distributions, move with the bulk velocity ${\mathbf{u_0}} = - u_0 \mathbf{\hat{x}}$ toward a reflecting wall at the leftmost boundary ($x = 0$), and a shock forms and propagates toward the $+\mathbf{\hat{x}}$ direction. 
Hence, simulations are performed in the rest frame of the shock downstream flow.
For the given preshock ion temperature, $T_i$, the flow Mach number, $M_0$, is related to the upstream bulk velocity as
\begin{equation}
\label{eq:e1}
M_0 \equiv \frac{u_0}{c_{\rm s1}} = \frac{u_0}{\sqrt{2\Gamma k_{B}T_{i1}/m_{i}}},
\end{equation}
where $c_{\rm s1}$ is the sound speed in the upstream medium and $\Gamma = 5/3$ is the adiabatic index. 
Thermal equilibrium is assumed for incoming plasmas, i.e., $T_{i1}=T_{e1}$, where $T_{e1}$ is the preshock electron temperature.
In typical PIC simulations, because of severe requirements for computational resources, reduced ion-to-electron mass ratios, $m_i/m_e < 1836$, are assumed.
Here, {\color{black} we consider the mass ratio of $m_i/m_e = 100$ and $400$}; electrons have the rest mass of $m_e=511~{\rm keV}/c^2$, while ``ions'' have reduced masses emulating the proton population.
In the limit of high $\beta$, the upstream flow speed in the shock rest frame can be expressed as $u_{\rm sh}\approx  u_0 \cdot r/(r-1)$,
where $r =  (\Gamma + 1)/(\Gamma - 1 + 2/M^2_{s})$ is the shock compression ratio, and the sonic Mach number, $M_{\rm s}$, of the induced shock is given as
\begin{equation}
\label{eq:e2}
M_{\rm s} \equiv \frac{u_{\rm sh}}{c_{\rm s1}} \approx M_0\frac{r}{r-1}.
\end{equation}

The magnetic field carried by incoming plasmas, $\mathbf{B_0}$, lies in the $x$-$y$ plane and the angle between $\mathbf{B_0}$ and the shock normal direction is the obliquity angle $\theta_{\rm Bn}$, as defined in the Introduction.
The initial electric field in the flow frame is zero everywhere, but the motional electric field, $\mathbf{E_0} = -\mathbf{u_0}/c \times \mathbf{B_0}$, is induced along $+\mathbf{\hat{z}}$ direction, where $c$ is the speed of light.
The strength of ${\mathbf{B_0}}$ is parameterized by $\beta$ as
\begin{equation}
\label{eq:e3}
\beta = \frac{8\pi nk_B(T_{i1} + T_{e1})}{B_0^2} = \frac{2}{\Gamma}\frac{M_{\rm A}^2}{M_{\rm s}^2},
\end{equation}
where $M_{\rm A}\equiv u_{\rm sh}/u_{\rm A}$ is the Alfv\'en Mach number of the shock. Here, $u_{\rm A}= B_0/\sqrt{4\pi nm_i}$ is the Alfv\'en speed, and $n = n_i = n_e$ are the number densities of incoming ions and electrons.
We consider $\beta=50$ and $100$, along with $k_BT_1 = k_BT_{i1} = k_BT_{e1} = 0.0168m_ec^2 = 8.6$ keV (or $T_{i1} = T_{e1} = 10^8$ K), relevant for typical ICM plasmas \citep{ryu2008, porter2015}.

{\color{black} The fast Mach number of MHD shocks is defined as $M_{\rm f} \equiv u_{\rm sh}/u_{\rm f}$, where the fast wave speed is
$u_{\rm f}^2=  \{ (c_{\rm s1}^2 + u_{\rm A}^2) + [ (c_{\rm s1}^2 + u_{\rm A}^2)^2 - 4 c_{\rm s1}^2 u_{\rm A}^2 \cos^2 \theta_{\rm Bn}]^{1/2} \}/2$.
In the limit of high $\beta$ (i.e., $c_{\rm s1}\gg u_{\rm A}$), $M_{\rm f}\approx M_{\rm s}$.}

The model parameters of our simulations are summarized in Table \ref{tab:t1}.
We adopt $\beta=100$, $\theta_{\rm Bn}=63^{\circ}$, and $m_i/m_e=100$ as the fiducial values of the parameters.
The incident flow velocity, $u_0$, is specified to induce shocks with $M_{\rm s} \approx 2 - 3$, which are characteristic for cluster merger shocks \citep[e.g.,][]{ha2018a}, as noted in the Introduction.
Models with different $M_{\rm s}$ are named with the combination of the letter `M' and sonic Mach numbers (for example, the M2.0 model has $M_{\rm s}=2.0$). 
Models with parameters different from the fiducial values have the names that are appended by a character for the specific parameter and its value.
For example, the M2.3-$\theta$73 model has $\theta_{\rm Bn}=73^{\circ}$, while the M2.3-m400 model has $m_i/m_e=400$.

Simulations are presented in units of the plasma skin depth, $c/ w_{\rm pe}$, and the electron plasma oscillation period, $w_{\rm pe}^{-1}$, where $w_{\rm pe} = \sqrt{4\pi e^{2}n/m_e}$ is the electron plasma frequency.
The $L_x$ and $L_y$ columns of Table \ref{tab:t1} denote the $x$ and $y-$sizes of the computational domain.
Except for the M3.0 model (see below), the longitudinal and transverse lengths are $L_x=7\times 10^3 c/w_{\rm pe}$ and $L_y=80 c/w_{\rm pe}$, respectively, which are represented by a grid of cells with size $\Delta x =\Delta y = 0.1 c/w_{\rm pe}$.
The last two columns show the end times of simulations in units of $w_{\rm pe}^{-1}$ and the ion gyration period, $\Omega_{\rm ci}^{-1}$, where $\Omega_{\rm ci} = eB_0/m_ic$ is the ion gyrofrequency.
The ratio of the two periods scales as $w_{\rm pe}/\Omega_{\rm ci} \propto (m_i/m_e)\sqrt{\beta}$.
For most models, simulations run up to $t_{\rm end} w_{\rm pe}\approx 1.13\times 10^5$, which corresponds to $t_{\rm end} \Omega_{\rm ci} \approx 30$ for $\beta=100$ and $m_i/m_e=100$.
The M3.0 model extends twice longer up to $t_{\rm end} w_{\rm pe}\approx 2.26\times 10^5$ or $t_{\rm end} \Omega_{\rm ci} \approx 60$, and correspondingly has a longer longitudinal dimension of $L_x=1.2\times 10^4 c/w_{\rm pe}$.
Comparison models with smaller $\beta$, M2.0-$\beta50$, M2.30-$\beta50$, and M3.0-$\beta50$, also go up to $t_{\rm end} \Omega_{\rm ci} \approx 30$ ($t_{\rm end} w_{\rm pe}\approx 8.0\times 10^4$).
{\color{black} The models with $m_i/m_e=400$, on the other hand, are calculated only up to $t_{\rm end} \Omega_{\rm ci} \approx 10$ ($t_{\rm end} w_{\rm pe}\approx 1.5\times 10^5$).}
Models with different $\Delta x / (c/w_{\rm pe})$, M2.3-r2 and M2.3-r0.5, are also considered to inspect the effects of spatial resolution.
In each cell, 32 particles (16 per species) are placed.
The time step is $\Delta t = 0.045 [w_{\rm pe}^{-1}]$.

Compared to the reference model reported by GSN14a and GSN14b, our fiducial models have higher $\beta$ (100 versus 20) and lower $T_{i1} = T_{e1}$ ($10^8$ K versus $10^9$ K).
As a result, our simulations run for a longer time, for instance, $\omega_{\rm pe}t_{\rm end}\approx 1.13\times 10^5$ to reach $t_{\rm end}\Omega_{\rm ci} \approx 30$.
And our shocks are less relativistic.
More importantly, this work also includes weaker shock models with $M_{\rm s} < 3.0$, while GSN14a and GSN14b considered only shocks with $M_{\rm s}=3.0$.

\section{Physics of $Q_\perp$-shocks}
\label{sec:s3}

\subsection{Critical Mach Numbers}
\label{sec:s3.1}

The structures and time variations of collisionless shocks are primarily governed by the dynamics of reflected ions and the waves excited by the relative drift between reflected and incoming ions.
In theories of collisionless shocks, hence, a number of {\it critical shock Mach numbers} have been introduced to describe ion reflection and upstream wave generation \citep[see][for a review]{balogh13}.
Although the main focus of this paper is the electron acceleration at $Q_\perp$-shocks, we here present a brief review on the ``shock criticalities'' due to reflected ions.

The reflection of ions has been often linked to the ``first critical Mach number'', $M_{\rm f}^*(\beta,\theta_{\rm Bn})$;
{\color{black} it was found for $u_{\rm n2}=c_{\rm s2}$ by applying the Rankine–Hugoniot jump relation to fast MHD shocks},
i.e., the condition that the downstream flow speed normal to the shock surface equals the downstream sound speed \citep[e.g.,][]{edmiston1984}. 
In {\it supercritical} shocks with $M_{\rm f} > M_{\rm f}^*$, the shock kinetic energy can not be dissipated enough through resistivity
and wave dispersion, and hence a substantial fraction of incoming ions should be reflected upstream in order to sustain the shock transition from the upstream to the downstream.
In {\it subcritcal} shocks below $M_{\rm f}^*$, on the other hand, the resistivity alone can provide enough dissipation to support a stable shock structure.
In collisionless shocks, however, the reflection of ions occurs at the shock ramp due to the magnetic deflection and the cross shock potential drop, the physics beyond the fluid description.
Hence, it should be investigated with simulations resolving kinetic processes.

In $Q_\parallel$-shocks, the first critical Mach number also denotes the minimum Mach number, above which kinetic processes trigger overshoot-undershoot oscillations in the density and magnetic field, and the shock structures may become non-stationary under certain conditions.
The reflection of ions is mostly due to the deceleration by the shock potential drop, and resonant and nonresonant waves are excited via streaming instabilities induced by reflected ions \citep[e.g.,][]{caprioli2014a,caprioli2014b}.
Such processes depend on shock parameters.
For instance, in shocks with higher $M_{\rm s}$ and hence higher shock kinetic energies, the structures tend to more easily fluctuate and become unsteady.
In high-$\beta$ plasmas, on the other hand, shocks could be stabilized against certain instabilities owing to fast thermal motions, which can subdue the relative drift between reflected and incoming particles;
thus, theoretical analyses based on the cold plasma assumption could be modified in high-$\beta$ plasmas.
In Paper I, we found that $M_{\rm f}^*\approx 2.3$ for $Q_\parallel$-shocks in ICM plasmas with $\beta\approx 100$, which is higher than the fluid prediction by \citet{edmiston1984}.
In $Q_\parallel$-shocks, the kinetic processes involved in determining $M_{\rm f}^*$ are also parts of the preacceleration of ions and hence the injection to the Fermi-I process of DSA.

In $Q_\perp$-shocks, both ions and electrons are reflected through the magnetic deflection;
the two populations are subject to deceleration by the magnetic mirror force due to converged magnetic field lines at the shock transition. 
In addition, the shock potential drop decelerates incident ions while it accelerates electrons toward the downstream direction.
The reflected particles gain energy through the gradient drift along the motional electric field at the shock surface (SDA).
Most of reflected ions, however, are trapped mostly at the shock foot before they advect downstream with the background magnetic field after about one gyromotion.
As a result, streaming instabilities are not induced in the upstream, and hence the ensuing CR proton acceleration is ineffective, as previously reported with hybrid and PIC simulations \citep[e.g.,][and Paper I]{caprioli2014a,caprioli2014b}.
However, still the dynamics of reflected ions is primarily responsible for the main features of the transition zone of $Q_\perp$-shocks \citep[e.g.,][]{treumann2008,treumann2009}.
For instance, the current due to the drift motion of reflected ions generates the magnetic foot, ramp, and overshoot.
And the charge separation due to reflected ions generates the ambipolar electric shock potential drop at the shock ramp.


{\color{black} In $Q_\perp$-shocks, the accumulation of reflected ions at the upstream edge of the foot may lead to the cyclic self-reformation of shock structures over ion gyroperiods and result in the excitation of low-frequency whistler waves in the shock foot region \citep[e.g.,][]{matsukiyo06,scholer2007}.}
This leads to the so-called ``second or whistler critical Mach number'', $M_{\rm w}^* \approx (1/2) \sqrt{m_i/m_e} \cos \theta_{\rm Bn}$ in the $\beta\ll 1$ limit \citep{kennel85,krasnoselskikh2002}.
In subcritical shocks with $M_{\rm f}<M_{\rm w}^*$,  linear whistler waves can phase-stand in the shock foot upstream of the ramp.
Dispersive whistler waves were found far upstream in interplanetary, subcritical shocks \citep[e.g.,][]{oka2006}.
Those waves interact with the upstream flow and contribute to the energy dissipation, effectively suppressing the shock reformation.
Above $M_{\rm w}^*$, stationary linear wave trains cannot stand in the region ahead of the shock ramp.

The ``third or nonlinear whistler critical Mach number'', $M_{\rm nw}^* \approx  \sqrt{m_i/2m_e} \cos \theta_{\rm Bn}$ in the $\beta\ll 1$ limit, was introduced to describe the non-stationarity of shock structures.
\citet{krasnoselskikh2002} predicted that in supercritical shocks with $M_{\rm f} > M_{\rm nw}^*$,
nonlinear whistler waves turn over because of the gradient catastrophe, leading to the non-stationarity of the shock front and quasi-periodic shock-reformation \citep[see][]{scholer2007}. 
However, \citet{hellinger2007} showed through 2D hybrid and PIC simulations that phase-standing oblique whistlers can be emitted 
in the foot even in supercritical $Q_\perp$-shocks, so the shock-reformation is suppressed, in 2D.
{\color{black} In fact, the nonstationarity and self-reformation of shock structures are an important long-standing problem in the study of collisionless shocks, 
which has yet to be fully understood \citep[e.g.,][]{scholer2003,lembege2004,matsukiyo06}.}

In the $\beta\ll 1$ limit (i.e., in cold plasmas), $M_{\rm w}^*=2.3$ and $M_{\rm nw}^*=3.2$ for the fiducial parameter values adopted for our PIC simulations ($m_i/m_e=100$ and $\theta_{\rm Bn}=63^{\circ}$). {\color{black} Hence, in some of the models considered here, $M_{\rm w}^*  < M_{\rm s} < M_{\rm nw}^*$, so the whistler waves induced by reflected ions could be 
be confined within the shock foot without overturning.}
However, these critical Mach numbers increase to $M_{\rm w}^*=9.7$ and $M_{\rm nw}^*=13.8$ for the true ratio of $m_i/m_e=1836$.
So in the ICM, weak $Q_\perp$-shocks are expected to be subcritical with respect to the two whistler critical Mach numbers, and so they would not be subject to self-reformation.
The confirmation of these critical Mach numbers, or improved estimations for $\beta \gtrsim 1$, through numerical simulations is very challenging, as noted above.
{\color{black} The excitation of oblique whistler waves and the suppression of shock-reformation via surface ripping require at least 2D simulations \citep[e.g.,][]{lembege2002,burgess2006}.}
The additional degree of freedom in higher dimensional simulations tends to stabilize some instabilities revealed in lower dimensional simulations.
Moreover, simulation results are often dependent on $m_i/m_e$, and the magnetic field configuration, i.e., whether ${\mathbf{B_0}}$ is in-plane or off-plane \citep{lembege2009}.
And adopting the realistic ratio of $m_i/m_e$ in PIC simulations is computationally very expensive, as pointed in the previous section.

As mentioned in the Introduction, GSN14a and GSN14b showed that in high-$\beta$, $Q_\perp$-shocks, electrons can be preaccelerated via multiple cycles of SDA due to the scattering by the upstream waves excited via the EFI.
We here additionally introduce the ``EFI critical Mach number'', $M_{\rm ef}^*$, above which the electron preacceleration is effective.
We seek it in the next section along with the relevant kinetic processes involved.

The space physics and ISM communities have been mainly interested in shocks in low-$\beta$ plasmas ($\beta \lesssim 1$), and hence the analytic relations simplified for cold plasmas are often quoted (e.g., the dispersion relation for fast magnetosonic waves used by \citet{krasnoselskikh2002}). 
In such works, $M_{\rm A}$ is commonly used to characterize shocks.
However, in hot ICM plasmas, shocks have $M_{\rm s} \approx M_{\rm f} \ll M_{\rm A}$, and magnetic fields play dynamically less important roles.
Moreover, the ion reflection at the shock ramp is governed mainly by $M_{\rm s}$ rather than $M_{\rm A}$ (e.g., Paper I).
Thus, in the rest of this paper, we will use the sonic Mach number $M_{\rm s}$ to characterize shocks.

\subsection{Energization of Electrons}
\label{sec:s3.2}

As mentioned in the Introduction, GSN14a discussed the relativistic SDA theory for electrons in $Q_\perp$-shocks, which involves the electron reflection at the shock and the energy gain due to the drift along the motional electric field.
In a subsequent paper, GSN14b showed that the electrons can induce the EFI, which leads to the excitation of oblique waves.
The electrons return back to the shock due to the scattering by those self-excited upstream waves and are further accelerated through multiple cycles of SDA (Fermi-like process).
Below, we follow these previous papers to discuss how the physical processes depend on the parameters such as $M_{\rm s}$, $\theta_{\rm Bn}$, and $T_1$ for the shocks considered here (Table \ref{tab:t1}).
Inevitably, we cite below some of equations presented in GSN14a and GSN14b.

\begin{figure*}[t]
\vskip -0.9cm
\hskip -0.3cm
\centerline{\includegraphics[width=1.1\textwidth]{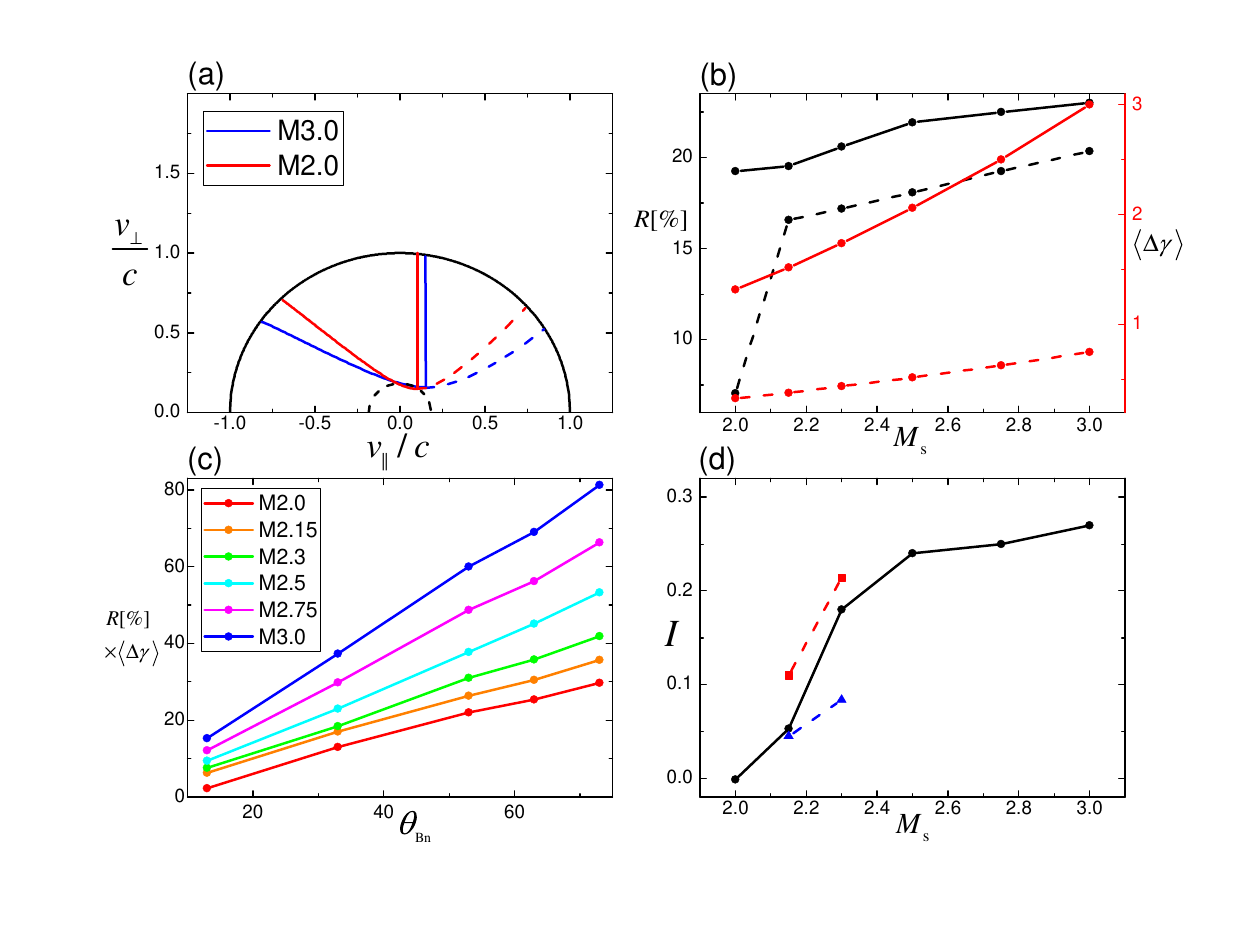}}
\vskip -1.5cm
\caption{\label{fig:f1} 
(a) Velocity diagram to analyze the electron reflection in weak ICM shocks; $v_{\parallel}$ and $v_{\perp}$ are the electron velocity components, parallel and perpendicular to the background magnetic field, respectively, in the upstream rest frame.
The black solid half-circle shows $v=c$, while the black dashed half-circle shows $v=v_{\rm th,e}$.
The red (for the M2.0 model with $M_{\rm s} = 2$ and $\theta_{\rm Bn} = 63^{\circ}$) and blue (for the M3.0 model with $M_{\rm s} = 3$ and $\theta_{\rm Bn} = 63^{\circ}$) vertical lines draw the reflection condition for $v_{\parallel}$ in Equation (\ref{vpar}), while the red and blue solid curves left to the vertical lines draw the reflection condition for  $v_{\perp}$ in Equation (\ref{vperp}).
The red and blue dashed curves right to the vertical lines draw the post-reflection velocity given in Equations (25)-(26) of GSN14a with the boundary values for the pre-reflection velocity given in Equations (\ref{vpar})-(\ref{vperp}) of this paper.
Electrons located in the region bounded by the colored vertical and solid lines are reflected to the region right to the vertical lines bounded by the dashed lines.
(b) The fraction of reflected electrons, $R$ in percentage (black), and the average energy gain via a single SDA, $\langle \Delta \gamma \rangle$ in units of $m_ec^2/k_BT$ (red), as a function of $M_{\rm s}$.
The solid lines are for $Q_\perp$-shocks with $\theta_{\rm Bn}=63^{\circ}$, while the dashed lines are for $Q_\parallel$-shocks with  $\theta_{\rm Bn}=13^{\circ}$.
(c) $R \cdot \langle \Delta \gamma \rangle$ as a function of $\theta_{\rm Bn}$ for different $M_{\rm s}$. 
(d) The EFI parameter, $I$, in Equation (\ref{firehose}) as a function of $M_{\rm s}$ for models with $\theta_{\rm Bn} = 63^{\circ}$ (black circles). 
{\color{black} The red squares are for models with $\theta_{\rm Bn} = 73^{\circ}$, while the blue triangles are for $\theta_{\rm Bn} = 53^{\circ}$}.
The instability condition is $I>0$.}
\end{figure*}

\subsubsection{Shock Drift Acceleration}
\label{sec:s3.2.1}

GSN14a derived the criteria for electron reflection by considering the dynamics of electrons in the so-called de Hoffmann–Teller (HT, hereafter) frame, in which the flow velocity is parallel to the background magnetic field and hence the motional electric field disappears both upstream and downstream of the shock \citep{dehoffmann1950}.
In the HT frame, the upstream flow has $u_t = u_{\rm sh} \sec \theta_{\rm Bn}$ along the background magnetic field.
Hereafter, $v_{\parallel}$ and $v_{\perp}$ represent the velocity components of incoming electrons, parallel and perpendicular to the background magnetic field, respectively,  in the upstream rest frame, and $\gamma_t \equiv (1-u_t^2/c^2)^{-1/2}$ is the Lorentz factor of the upstream flow in the HT frame.

The reflection criteria can be written as
\begin{equation}
\label{vpar}
v_{\parallel} < u_t
\end{equation}
(Equation (19) of GSN14a), and
\begin{eqnarray}
\label{vperp}
v_{\perp} & \gtrsim & \gamma_t \tan \alpha_0 \cdot 
\big[(v_{\parallel} - u_t)^2 + 2 c^2 \cos^2\alpha_0 \Delta \phi \cdot G \cdot F \nonumber\\
& + & \{c^2 \cos^2\alpha_0 \cdot G - (v_{\parallel} - u_t)^2\}\Delta \phi^2\big]^{1/2},
\end{eqnarray}
assuming that the normalized cross-shock potential drop is $\Delta \phi(x)\equiv e [\phi^{\rm HT}(x) -\phi^{\rm HT}_0]/m_ec^2 \ll 1$.
Here, $G\equiv (1-v_{\parallel}u_t/c^2)^2$, $F\equiv [1- (v_{\parallel} - u_t)^2/(G c^2 \cos^2\alpha_0)]^{1/2}$, and $\alpha_0 \equiv \sin^{-1} (1/\sqrt{b})$ with the magnetic compression ratio $b \equiv B(x)^{\rm HT}/ B_0^{\rm HT}$.
The superscript HT denotes the quantities in the HT frame.
Note that for $\Delta \phi(x)=0$, Equation (\ref{vperp}) becomes the same as Equation (20) of GSN14a.

In Figure \ref{fig:f1}(a), the red and blue solid lines mark the boundaries of the reflection criteria in Equations (\ref{vpar}) and (\ref{vperp}) for the M2.0 and M3.0 models, respectively.
The red and blue dashed curves right to the vertical lines are the post-reflection velocities calculated with Equations (25) and (26) of GSN14a by inserting the boundary values of Equations (\ref{vpar}) and (\ref{vperp}).
For $b$ and $\Delta \phi$, the values estimated at the shock surface from simulation data were used.
The solid black half-circle shows $v \equiv (v_{\parallel}^2 + v_{\perp}^2)^{1/2} = c$, while the dashed black half-circle shows $v = v_{\rm th,e}$, where $v_{\rm th,e} = \sqrt{2k_B T_{e1}/m_e}$ is the electron thermal speed of the incoming flow.

As in GSN14b, we estimated {\it semi-analytically} the amount of the incoming electrons that satisfy the reflection condition, that is, those bounded by the colored solid curves and the colored vertical lines together with the black circle in Figure \ref{fig:f1}(a).
In Figure \ref{fig:f1}(b), the fraction of the reflected electrons, $R$, estimated for $Q_\perp$-shocks with $\theta_{\rm Bn}=63^{\circ}$ is shown by the black filled circles connected with the black solid line, while $R$ for $Q_\parallel$-shocks with $\theta_{\rm Bn}=13^{\circ}$ is shown by the black filled circles connected with the black dashed line.
In $Q_\perp$-shocks, the reflection fraction, $R$, is quite high and increases with $M_{\rm s}$, ranging $\sim 20-25$ \% for $2 \le M_{\rm s} \le 3$.
In $Q_\parallel$-shocks, $R$ is also high, ranging $\sim 17-20$ \%, for $2.15 \le M_{\rm s} \le 3$, but drops sharply at $M_{\rm s} = 2$.

We point out that the electron reflection becomes ineffective for superluminal shocks with large obliquity angles (i.e., $u_{\rm sh}/\cos \theta_{\rm Bn} \ge c$), 
since the electrons streaming upstream along the background field cannot outrun the shocks (see GSN14b).
The obliquity angle for the superluminal behavior is $\theta_{\rm sl} \equiv \arccos(u_{\rm sh}/c) = 86^{\circ}$ for the shock in M3.0 with $m_i/m_e = 100$ 
and $T_1 = 10^8$ K, and it is larger for smaller $M_{\rm s}$.
This angle is larger than $\theta_{\rm Bn}$ of our models in Table \ref{tab:t1}, and hence all the shocks considered here are subluminal.

For given $T_1$ and $M_{\rm s}$, the reflection of electrons is basically determined by $b(x)$ and $\Delta \phi(x)$, 
which quantify the magnetic deflection and the acceleration at the shock potential drop.
Both $b(x)$ and $\Delta \phi(x)$ increase with increasing $\theta_{\rm Bn}$.
{\color{black} Larger $b$ enhances the electron reflection (positive effect), while larger $\Delta \phi(x)$ suppresses it (negative effect).
In GSN14b, shocks are semi-relativistic with $\Delta \phi\sim 0.1-0.5$, and hence the negative effect of the potential drop is substantial.}
However, in our models, shocks are less relativistic because of the lower temperature adopted, and $\Delta \phi \sim {m_i u_{\rm sh}^2}/2m_e c^2 \ll 1$.
As a result, the magnetic deflection dominates over the acceleration by the cross-shock potential, leading to higher $R$ at higher $\theta_{\rm Bn}$.
{\color{black} GSN14b showed that SDA becomes inefficient for $u_t \gtrsim v_{\rm th,e}$ ($\cos\theta_{\rm Bn} \lesssim \cos\theta_{\rm limit} = M_{\rm s} \sqrt{m_e/m_i}$),
which is more stringent than the superluminal condition ($u_t > c$).
So the electron reflection fraction begins to decrease for $\theta_{\rm Bn} \gtrsim 60^{\circ}$ in their models. 
Although not shown here, in our models, $R$ monotonically increases with the obliquity angle for a given $M_{\rm s}$,
because the adopted $\theta_{\rm Bn}$ ($\le 73^{\circ}$) is smaller than the limiting obliquity angle, $\theta_{\rm limit}$, for $M_{\rm s}=2-3$ and $m_i/m_e=100$.}

The reflected electrons gain the energy via SDA.
We estimated the energy gain from a single SDA cycle as
\begin{equation}
\label{delgamma}
\Delta \gamma \equiv \gamma_r - \gamma_i = \frac{2u_t(u_t-v_{\parallel})}{c^2 -u_t^2} \gamma_i,
\end{equation}
where $\gamma_i$ and $\gamma_r$ are the Lorentz factors for the pre-reflection and post-reflection electron velocities, respectively (Equation (24) of GSN14a).
For given $T_1$ (or given $c_{\rm s1}$), $u_t$ and $\Delta \gamma$ depend on $M_{\rm s}$ and $\theta_{\rm Bn}$.
For the shocks considered here, $\gamma_i\approx 1$ and $u_t \ll c$, so $\Delta \gamma \approx 2[(u_t/c)^2 - u_t v_{\parallel}/c^2]$.
In Figure \ref{fig:f1}(b), the red filled circles connected with the red solid line show the average energy gain, $\langle \Delta \gamma \rangle$ in units of $m_ec^2/k_BT_e$, estimated for $Q_\perp$-shocks with $\theta_{\rm Bn}=63^{\circ}$.
The red filled circles with the red dashed line show the quantity for $Q_\parallel$-shocks with $\theta_{\rm Bn}=13^{\circ}$. 
Here, the average was taken over the incoming electrons of Maxwellian distributions, 
so $\langle \Delta \gamma \rangle$ shown are the representative values during the initial development stage of suprathermal particles.

In addition, the product of $R$ and $\langle \Delta \gamma \rangle$ is plotted as a function of $\theta_{\rm Bn}$ for different $M_{\rm s}$ in Figure \ref{fig:f1}(c).
For the models in Table \ref{tab:t1}, $R$ was calculated using $b$ and $\Delta \phi$ estimated at the shock surface from simulation data, as mentioned above.
For the rest, the values of $b$ and $\Delta \phi$ for the models with $\theta_{\rm Bn} = 13^{\circ}$ presented in Paper I were adopted for $Q_\parallel$-shocks, while the values for the models with $\theta_{\rm Bn} = 63^{\circ}$ presented in this work were adopted for $Q_\perp$-shocks.
Figures \ref{fig:f1}(b)-(c) show that more electrons are reflected and higher energies are achieved at higher $M_{\rm s}$ and larger $\theta_{Bn}$. 

\begin{figure*}[t]
\vskip -1cm
\hskip -0.4cm
\centerline{\includegraphics[width=1.1\textwidth]{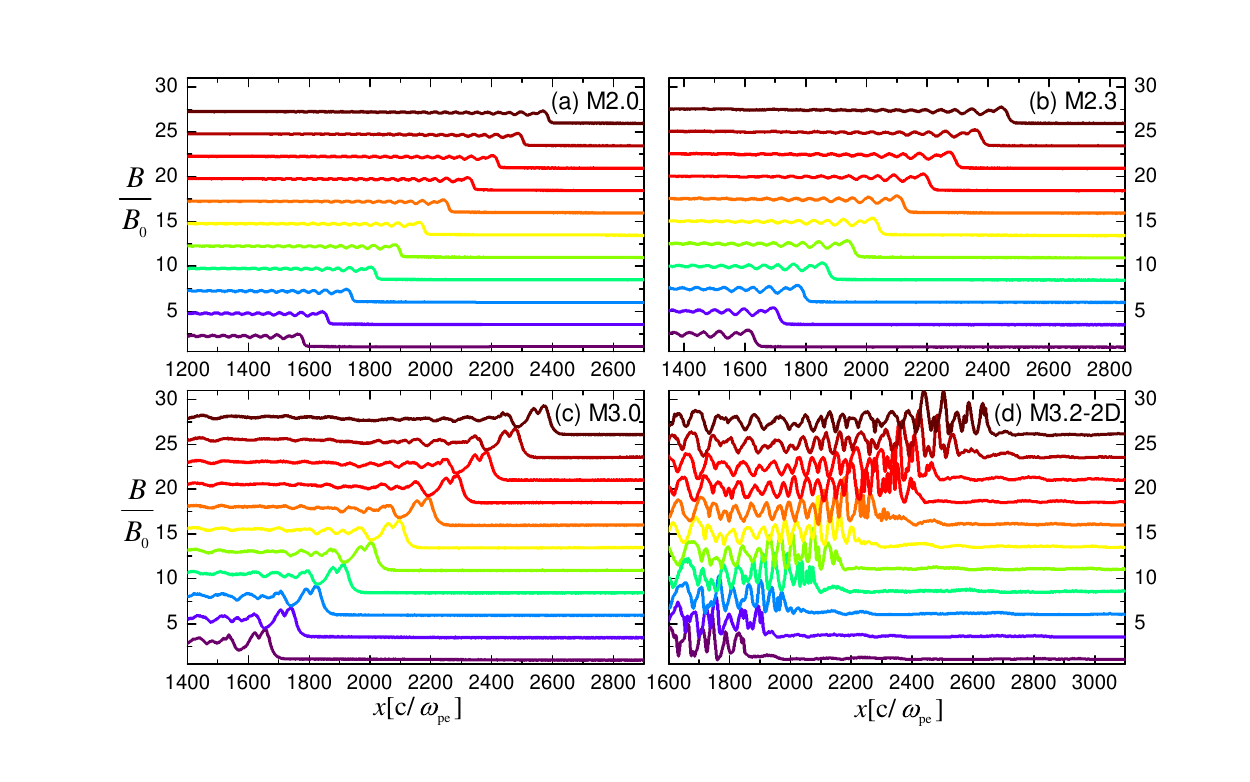}}
\vskip -0.9cm
\caption{\label{fig:f2}
Stack plots of the total magnetic field strength, averaged over the transverse direction, $B$, in the M2.0, M2.3, and M3.0 models from $t\Omega_{\rm ci}=20$ (bottom) to $t\Omega_{\rm ci}=30$ (top). The M3.2-2D model represents the $Q_\parallel$-shock with $M_{\rm s}=3.2$ and $\theta_{\rm Bn}=13^{\circ}$, taken from Paper I. Here, $B_0$ is the magnetic field strength far upstream.}
\end{figure*}

\begin{figure}[t]
\vskip 0cm
\hskip -0.2cm
\centerline{\includegraphics[width=0.55\textwidth]{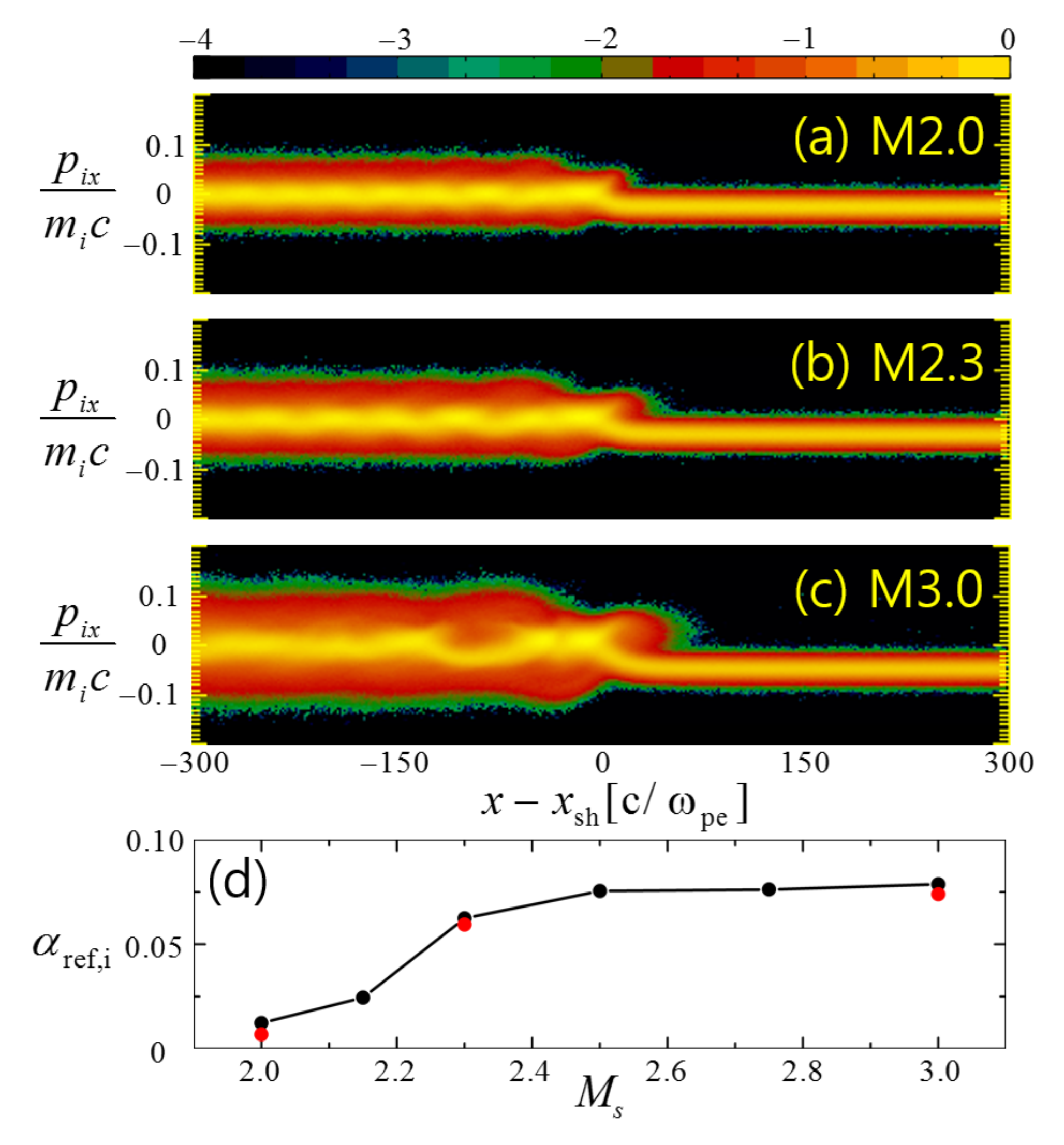}}
\vskip -0.3cm
\caption{\label{fig:f3}
Ion phase-space distributions in the $x-p_{ix}$ plane for the M2.0 model (a), the M2.3 model (b), and the M3.0 models (c) at $t\Omega_{\rm ci} \approx 30$. The $x$-coordinate is measured relative to the shock position, $x_{\rm sh}$, in units of $c/w_{\rm pe}$. The bar at the top displays the color scale for the log of the ion phase-space density (arbitrary units).
In panel (d), the black circles show
the fraction of reflected ions in the shock ramp region of $0\le x-x_{\rm sh} \le 60 c/w_{\rm pe}$ at $t\Omega_{\rm ci} \approx 30$ for the fiducial models with $m_i/m_e=100$, while the red circles show
the same fraction in $0\le x-x_{\rm sh} \le 240 c/w_{\rm pe}$  at $t\Omega_{\rm ci} \approx 10$ for the three models with $m_i/m_e=400$.}
\end{figure}

\subsubsection{Electron Firehose instability}
\label{sec:s3.2.2}

GSN14b performed periodic-box simulations with beams of streaming electrons in order to isolate and study the EFI due to the reflected and SDA-energized electrons.
They found the followings:
non-propagating ($\omega_r \approx 0$), oblique waves with wavelengths $\sim (10-20)c/w_{\rm pe}$ are excited dominantly,
$\delta B_z$ is stronger than $\delta B_x$ and $\delta B_y$ (the initial magnetic field is in the $x$-$y$ plane), and both the growth rate and the dominant wavelength of the instability {\color{black} are not sensitive to the mass ratio} $m_i/m_e$.
These results are consistent with the expectations from the previous investigations of oblique EFI \citep[e.g.,][]{gary2003}.

The EFI criterion in weakly magnetized plasmas can be defined as
\begin{equation}
\label{firehose}
I \equiv 1 - \frac{T_{e\perp}}{T_{e\parallel}} - \frac{1.27}{\beta_{e\parallel}^{0.95}} > 0,
\end{equation}
where $\beta_{e\parallel} \equiv 8\pi n_e k_B T_{e\parallel}/B_0^2$ is the electron beta parallel to the initial magnetic field (Equation (10) of GSN14b).
{\color{black} Equation (\ref{firehose}) indicates that the instability parameter, $I$, is larger for higher $\beta_{e\parallel}$ for a given value of
$T_{e\perp}/T_{e\parallel}$. For higher $M_{\rm s}$, $R$ is larger and $T_{e\perp}/T_{e\parallel}$ is smaller, leading to larger $I$.} 

Figure \ref{fig:f1}(d) shows the instability parameter of shocks with $\theta_{\rm Bn} = 63^{\circ}$, as a function of $M_{\rm s}$, estimated using the velocity distributions of the electrons which are located within $(0-1) r_{L,i}$ ($r_{L,i}$ is the ion Larmor radius with the upstream field $B_0$) upstream from the shock position in simulation data.
For the $M_{\rm s}=2.0$ model, $I \lesssim 0$ with almost no temperature anisotropy, so the upstream plasma should be stable against the EFI.
{\color{black} This finding, which will be further updated with simulation results in the next section, suggests that 
the preacceleration of electrons due to the EFI may not operate effectively in very weak shocks.
For $M_{\rm s} > 2$, on the other hand, the EFI criterion is satisfied and $I$ increases with increasing $M_{\rm s}$, 
implying that larger temperature anisotropies ($T_{e\parallel} > T_{e\perp}$) at higher $M_{\rm s}$ shocks induce stronger EFIs.
Also the figure indicates that $I$ increases steeply around $M_{\rm s}\approx 2.2-2.3$.
Additional data points marked with the blue triangles connected with the blue dashed line ($\theta_{\rm Bn} = 53^{\circ}$) 
and the red squares connected with the red dashed line ($\theta_{\rm Bn} = 73^{\circ}$) show that
$Q_\perp$-shocks with higher obliquity angles are more unstable to the EFI.}

\section{Results}
\label{sec:s4}

\subsection{Shock Structures}
\label{sec:s4.1}

As discussed in Section \ref{sec:s3.1}, the criticality defined by the first critical Mach, $M_{\rm f}^*$, primarily governs the structures and time variations of collisionless shocks.
In {\it subcritical} shocks, most of the shock kinetic energy is dissipated at the shock transition, resulting in relatively smooth and steady structures. 
In {\it supercritical} shocks, on the other hand, reflected ions induce overshoot-undershoot oscillations in the shock transition and ripples along the shock surface.
$Q_\parallel$-shocks with $M_{\rm f} > M_{\rm f}^*$ may undergo quasi-periodic reformation owing to the accumulation of upstream low-frequency waves.
$Q_\perp$-shocks are less prone to reformation, because reflected ions mostly advect downstream after about one gyromotion.

{\color{black} 
Figure \ref{fig:f2} compares the magnetic field structure for $Q_\perp$ and $Q_\parallel$-shocks with different $M_{\rm s}$.
In the $Q_\perp$-shocks, the overshoot-undershoot oscillation becomes increasingly more evident for higher $M_{\rm s}$,
but the shock structure seems to be quasi-stationary without any signs of reformation.
This is consistent with the fact that the nonlinear whistler critical Mach number for our fiducial models is $M_{\rm nw}^*=3.2$.
On the other hand, the $Q_\parallel$-shock in the M3.2-2D model exhibits quasi-periodic reformations.

According to the fluid description of \citet{edmiston1984}, $M_{\rm f}^* \approx 1$ for $Q_\perp$-shocks in high-$\beta$ plasmas,
so the fraction of reflected ions is expected be relatively high in all the shock models under consideration.
As can be seen in the phase-space distribution of protons in Figure \ref{fig:f3}(a)-(c), the back-streaming ions turn around mostly within about one ion gyroradius in the shock ramp ($x-x_{\rm sh} \lesssim 60 c/w_{\rm pe}$).
Note that with $m_i/m_e=100$, in the M3.0 model, the shock ramp corresponds to the region of $x-x_s < 60 c/\omega_{\rm pe}$, while the foot extends to $x-x_s\approx r_{L,i}\approx 200 c/\omega_{\rm pe}$ \citep[e.g.,][]{balogh13}.
In the M2.0 model, $r_{L,i}$ is smaller and so the characteristic widths of the ramp and foot are accordingly smaller.
Figure \ref{fig:f3}(d) shows that the ion reflection fraction, $\alpha_{\rm ref,i}= n_{\rm ref,i}/n_i$, increases with increasing $M_{\rm s}$, and such trend is almost independent of the mass ratio $m_i/m_e$.
Here, $n_{\rm ref,i}$ was calculated as the number density of ions with $v_x > 0$ in the shock rest frame in the ramp region.
Since $\alpha_{\rm ref,i}$ increases abruptly at $M_{\rm s}\approx 2.2-2.3$,
we may regard $M_{\rm f}^* \approx 2.3$ as an {\it effective} value for the first critical Mach number, above which high-$\beta$ $Q_\perp$-shocks reflect a sufficient amount of 
incoming ions and become supercritical.
From Figure \ref{fig:f2}, we can see that ensuing oscillations in shock structures appear noticeable in earnest only for $M_{\rm f}\gtrsim 2.3$.
Our estimation of $M_{\rm f}^*$ is higher than the prediction of \citet{edmiston1984}.  
This might be partly because in high-$\beta$ plasmas, kinetic processes due to fast thermal motions could suppress some of microinstabilities driven by the relative drift between backstreaming and incoming ions, as mentioned before.}

\begin{figure*}[t]
\vskip 0cm
\hskip 0cm
\centerline{\includegraphics[width=1\textwidth]{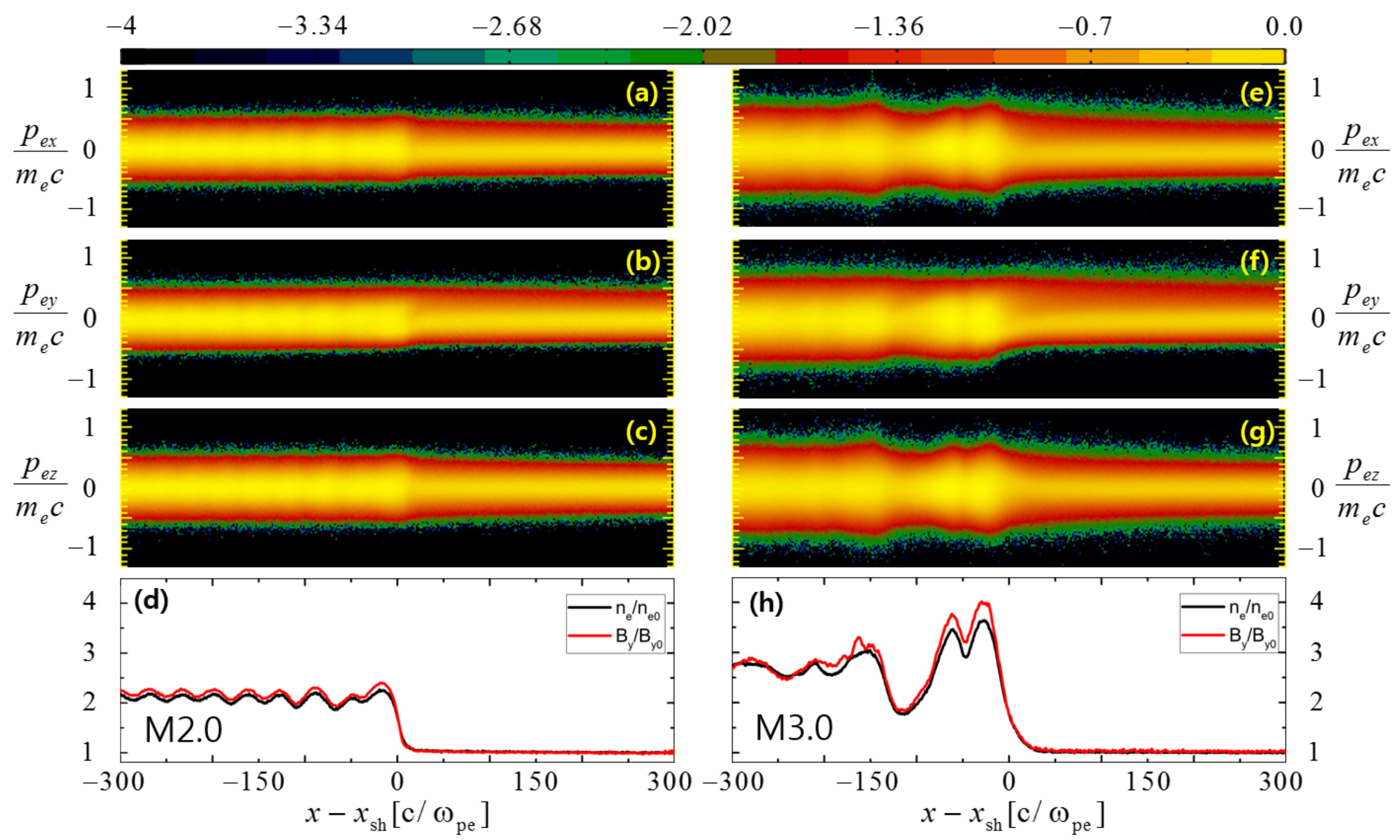}}
\vskip -0.1cm
\caption{\label{fig:f4}
Electron phase-space distributions and shock structures for the M2.0 model (left panels) and the M3.0 model (right panels) at $w_{\rm pe}t \approx 1.13\times 10^5$ ($t\Omega_{\rm ci} \approx 30$). The $x$-coordinate is measured relative to the shock position, $x_{sh}$, in units of $c/w_{\rm pe}$. 
From top to bottom, the distributions in $x-p_{\rm ex}$, $x-p_{\rm ey}$, and $x-p_{\rm ez}$, and the distributions of electron number density $n_e$ and transverse magnetic field $B_y$ in units of upstream values are shown.
The bar at the top displays the color scale for the log of the electron phase-space density (arbitrary units).}
\end{figure*}

\subsection{Electron Preacceleration}
\label{sec:s4.2}

\begin{figure*}[t]
\vskip -1.2cm
\hskip -0.3cm
\centerline{\includegraphics[width=1.2\textwidth]{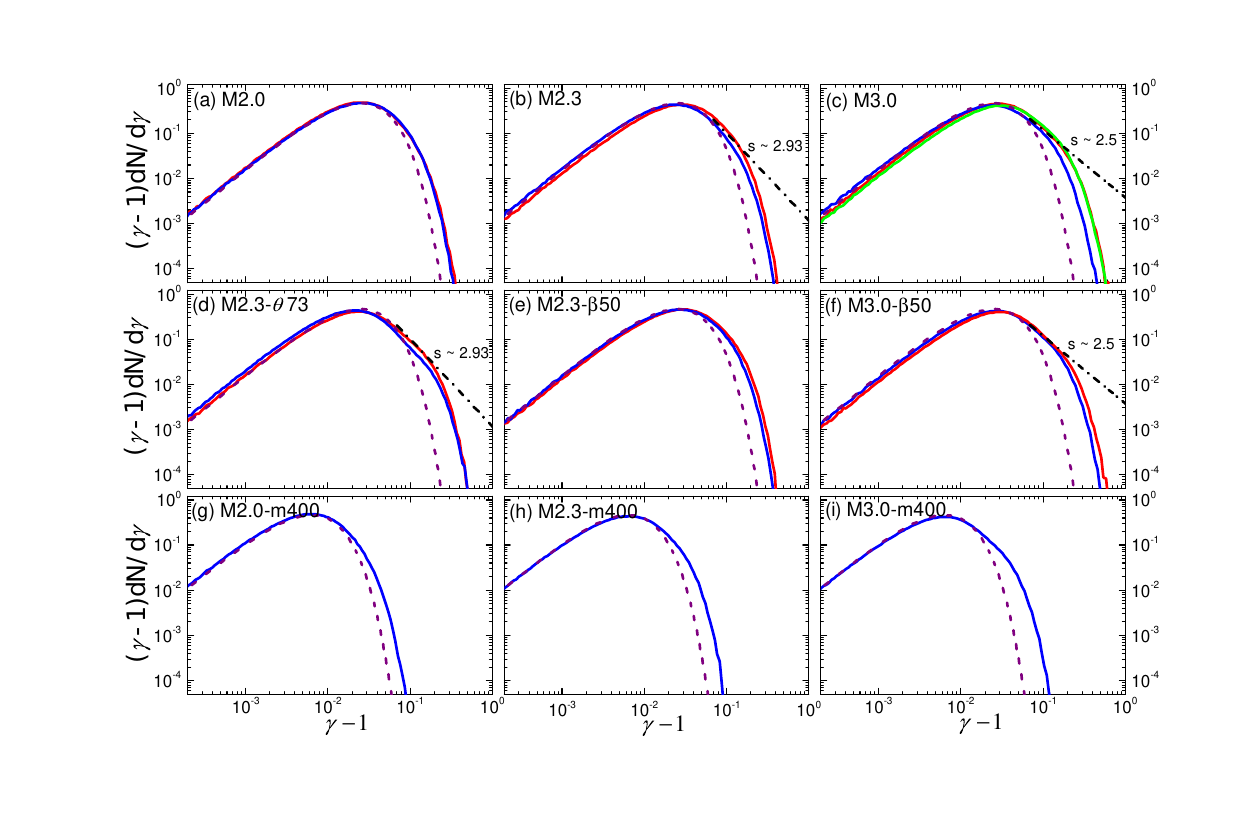}}
\vskip -1.4cm
\caption{\label{fig:f5}
Upstream electron energy spectra at $t\Omega_{\rm ci} = 10$ (blue lines), $t\Omega_{\rm ci} = 30$ (red), and $t\Omega_{\rm ci} = 60$ (green) in various models.
The spectra were taken from the region of $(0 - 1) r_{L,i}$ upstream of the shock.
The black dot-dashed lines indicate the test-particle power-laws of Equation (\ref{ngamma}), while the purple dashed lines show the Maxwellian distributions in the upstream region.}
\end{figure*}

Reflected electrons are energized via SDA at the shock ramp, and the consequence can be observed in the phase-space distribution of electrons in Figure \ref{fig:f4}, (a)-(c) for the M2.0 model and (e)-(g) for the M3.0 model.
Since ${\mathbf{B}_0}$ is in the $x-y$ plane, electrons at first gain the $z$-momentum, $p_{ez}$, through the drift along the motional electric field, $\mathbf{E_0} = -\mathbf{v_0}/c \times \mathbf{B}_0$, and then the gain is distributed to $p_{ex}$ and $p_{ey}$ during gyration motions.
In addition, reflected electrons, streaming along the background magnetic field with small pitch angles in the upstream region, have larger positive $p_y$ than $p_x$.
Figures \ref{fig:f4}(d) and (h) also show the distributions of electron density $n_e$ (black curve) and $B_y$ (red curve) around the shock transition.

If electrons are accelerated via the full Fermi-I process (i.e., DSA) and in the test-particle regime, the momentum distribution follows the so-called DSA power-law:
\begin{equation}
\label{eq:e4}
f(p) \approx f_N \left({p \over p_{\rm inj}}\right)^{-q} \exp \left[-\left({p \over p_{\rm max}}\right)^2\right],
\label{finj}
\end{equation}
where $f_N$ is the normalization factor and $q(M_{\rm s})=3r/(r-1)$ is the slope \citep{drury1983, kang2010}.
Here, $p_{\rm max}$ is the maximum momentum of accelerated electrons that increases with the shock age before any energy losses set in.
The injection momentum, $p_{\rm inj}$, is the minimum momentum with which electrons can diffuse across the shock and be injected to the full Fermi-I process as described in the Introduction. 
It marks roughly the boundary between the thermal and nonthermal momentum distributions.
The momentum spectrum in Equation (\ref{finj}) can be transformed to the energy spectrum in terms of the Lorentz factor as
\begin{equation}
\label{ngamma}
4\pi p^2f(p)\frac{dp}{dE} \propto \frac{dN}{d\gamma}\propto (\gamma-1 )^{-s},
\end{equation}
where the slope is $s(M_{\rm s}) = q(M_{\rm s})-2$. 
For instance, $s=2.5$ for $M_{\rm s}=3.0$, while $s=2.93$ for $M_{\rm s}=2.0$.

The injection momentum, which can be estimated as $p_{\rm inj} \sim 3 p_{\rm th,i}$ \citep[e.g.,][Paper I]{kang2002,caprioli2015}, is well beyond the highest momentum that electrons can achieve in our PIC simulations. 
In the M3.0 model, for example, $p_{\rm inj}$ corresponds to $\gamma_{\rm inj} \approx 10$, while electrons of highest momenta reach only $\gamma \lesssim 2$ (see Figure \ref{fig:f5}).
In other words, our simulations could follow only the preacceleration of {\it suprathermal} electrons, which are not energetic enough to diffuse across the shock.
Thus, the DSA slope, $s(M_{\rm s})$, is not necessarily reproduced in the energy spectra of electrons.
However, the development of power-law tails with $s(M_{\rm s})$ may indicate that the preaccelerated electrons have undergone a Fermi-like process, as proposed by GSN14a and GSN14b.

The upper panels of Figure \ref{fig:f5} compare the electron energy spectra, $(\gamma - 1) d N/ d \gamma$, taken from the upstream region of $(0 - 1) r_{L,i}$, ahead of the shock, at $t\Omega_{\rm ci} = 10$ (blue lines) and $30$ (red lines), in the models with different $M_{\rm s}$. 
In the case of the M3.0 model, the simulation is perform longer, and the spectrum at $t\Omega_{\rm ci} = 60 $ is also shown with the green line (which almost overlaps with the red line).
As described in Section \ref{sec:s3.1}, reflected electrons gain energy initially via SDA, and may continue to be accelerated via a Fermi-like process and multiple cycles of SDA, if oblique waves are excited by the EFI.
Two points are noticed:
(1) In the M2.0 model, the blue and red lines almost coincide, indicating almost no change of the spectrum from $t\Omega_{\rm ci} = 10$ to 30.
The spectrum is similar to that of the electrons energized by a single cycle of SDA, which was illustrated in Figure 7 of GSN14a.
So the Fermi-like process, followed by the EFI, may not efficiently operate in this model.
(2) The M3.0 model, on the other hand, exhibits a further energization from $t\Omega_{\rm ci} = 10$ to $30$, demonstrating the presence of a Fermi-like process.
However, there is no difference in the spectra of $t\Omega_{\rm ci} = 30$ and $60$.
As a matter of fact, the energy spectrum of suprathermal electrons seems to saturate beyond $t\Omega_{\rm ci}\approx 20$ (not shown in the figure).
This should be due to the saturation of the EFI and the lack of further developments of longer wavelength waves (see the next subsection for further discussions). 

{\color{black} The middle and lower panels of Figure \ref{fig:f5} show the electron energy spectra in models with different parameters.
The models with $m_i/m_e=400$ were followed only up to $t_{\rm end} \Omega_{\rm ci}=10$ (blue lines), 
because longer computing time is required for larger $m_i/m_e$.
Comparison of the two sets of models with different values of $m_i/m_e$
confirms that the EFI is {\color{black} almost} independent of $m_i/m_e$ for sufficiently large mass ratios, as previously shown by \citet{gary2003} and GSN14b, and so is the electron acceleration.
Figure \ref{fig:f5}(d) for the M2.3-$\theta$73 model indicates that SDA and hence the EFI is more efficient at higher obliquity angles, {\color{black} which is consistent with Figure \ref{fig:f1}(c).}
Figure \ref{fig:f5}(e) and (f) for the models with $\beta=50$ demonstrate that the EFI is more efficient at higher $\beta$.
All the models with $M_{\rm s}\gtrsim 2.3$ show marginal power-law-like tails beyond the spectra energized by a single cycle of SDA.}

With the M2.3-r2 and M2.3-r0.5 models, we examined how the electron energy spectrum depends on the grid resolution, although the comparison plots are not shown.
Our simulations with different $\Delta x$ produced essentially the same spectra, especially for the suprathermal part.

In Paper I, we calculated the injection fraction, $\xi(M_{\rm s},\theta_{\rm Bn}, \beta)$, of nonthermal protons with $p\ge p_{\rm inj}$ for $Q_{\parallel}$ shocks, as a measure of the DSA injection efficiency.
Since the simulations in this paper can follow only the preacceleration stage of electrons via an upstream Fermi-like process, we define and estimate the ``fraction of suprathermal electrons'' as follows:
\begin{equation}
\label{zeta}
\zeta \equiv \frac{1}{n_2}\int_{p_{\rm spt}}^{p_{\rm max}} 4\pi \langle f(p)\rangle p^2 dp , 
\end{equation}
where $\langle f(p)\rangle $ is the electron distribution function, averaged over the upstream region of $(0 - 1) r_{L,i}$, ahead of the shock.
For the ``suprathermal momentum'', above which the electron spectrum changes from Maxwellian to power-law-like distribution, we use $p_{\rm spt} \approx 3.3 p_{\rm th,e}$.
Note that $p_{\rm spt} \approx p_{\rm inj}(m_i/m_e)^{-1/2} $.
For the M3.0 model, for instance, $p_{\rm spt}$ corresponds to $\gamma \approx 1.25$.
Different choices of $p_{\rm spt}$ result in different values of $\zeta$, of course, but the dependence on the parameters such as $M_{\rm s}$ and $\theta_{\rm Bn}$, does not change much.

\begin{figure}[t]
\vskip -0.6cm
\hskip 0cm
\centerline{\includegraphics[width=0.65\textwidth]{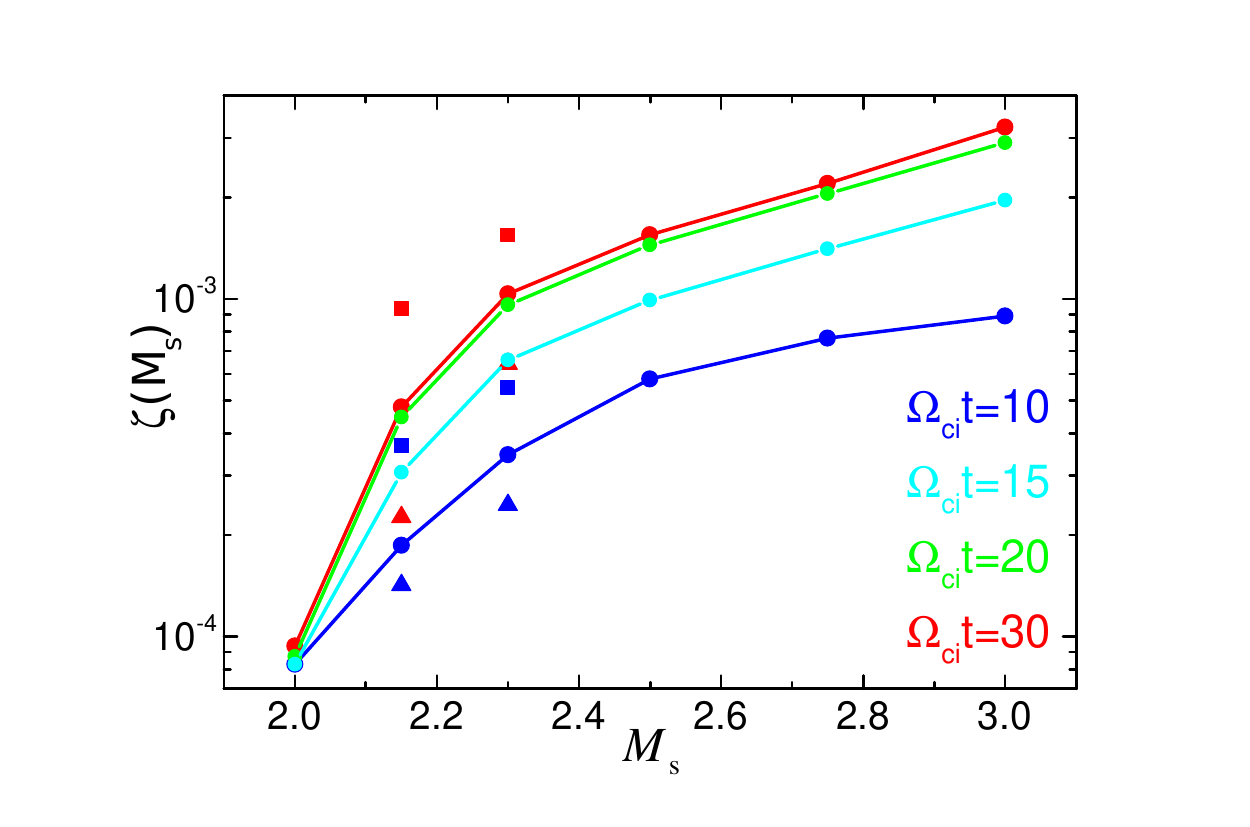}}
\vskip -0.6cm
\caption{\label{fig:f6}
Suprathremal fraction, $\zeta$, defined in Equation (\ref{zeta}), as a function of $M_{\rm s}$ for the fiducial models ($\theta_{\rm Bn} = 63^{\circ}$) at $t\Omega_{\rm ci} = 10$ (blue circles), 15 (cyan circles), 20 (green circles), and $t\Omega_{\rm ci} = 30$ (red circles). 
The triangles are for the models with $\theta_{\rm Bn}=53^{\circ}$ at $t\Omega_{\rm ci} = 10$ (blue) and 30 (red), 
while the squares are for the models with $\theta_{\rm Bn}=73^{\circ}$ at $t\Omega_{\rm ci} = 10$ (blue) and 30 (red).}
\end{figure}

{\color{black} 
In Figure \ref{fig:f6}, the circles connected with solid lines show the suprathermal fraction, $\zeta(M_{\rm s})$, 
for the fiducial models with $\theta_{\rm Bn} = 63^{\circ}$ at $t\Omega_{\rm ci} =10-30 $.
This fraction is expected to increase with increasing $M_{\rm s}$, since the EFI parameter, $I$, is larger for higher $M_{\rm s}$ (see Figure \ref{fig:f1}(d)).
Moreover, it increases with time until $t\Omega_{\rm ci} \approx 20$ due to a Fermi-like process,
as shown in Figure \ref{fig:f5}, except for the M2.0 model where the increase in time is insignificant.
However, $\zeta$ seems to stop growing for $t\Omega_{\rm ci} \gtrsim 20$, indicating the saturation of electron preacceleration.
This is related with the reduction of temperature anisotropy via electron scattering and the ensuing decay of EFI-induced waves, which will be discussed more in the next section.}

\begin{figure*}[t]
\vskip -0.6cm
\hskip -0.6cm
\centerline{\includegraphics[width=1.2\textwidth]{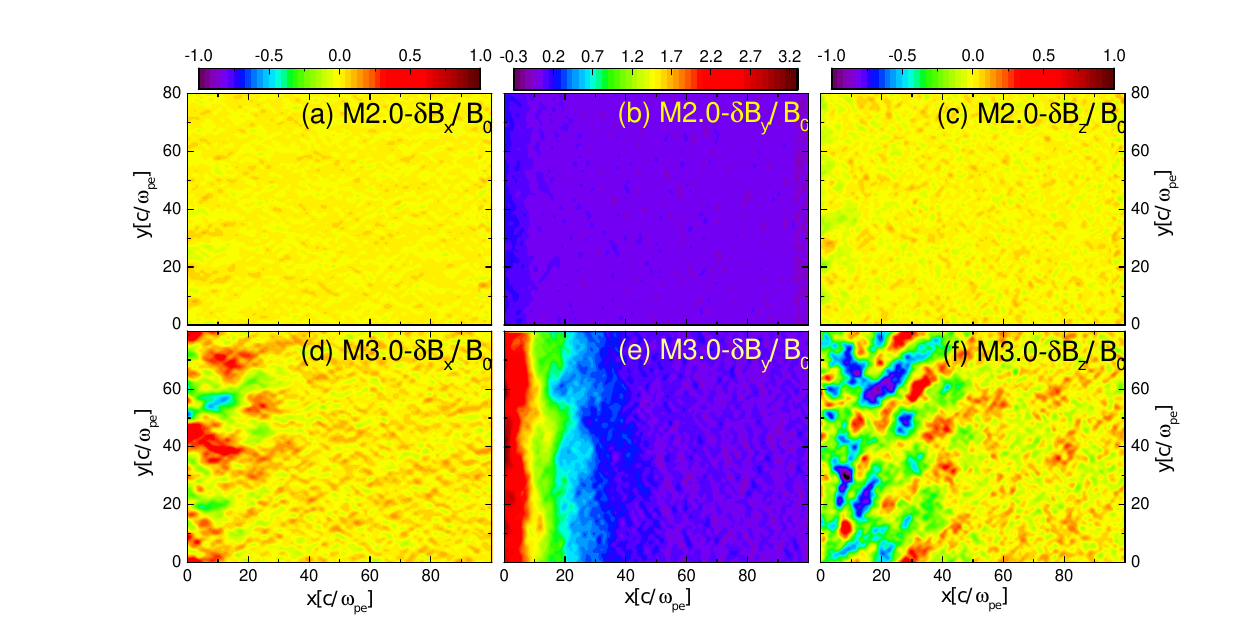}}
\vskip -0.3cm
\caption{\label{fig:f7}
Magnetic field fluctuations, $\delta B_x$ in (a) and (d), $\delta B_y$ in (b) and (e), and $\delta B_z$ in (c) and (f), normalized to $B_0$, in the upstream region of $0<(x-x_{\rm sh})w_{\rm pe}/c < 100$ at $w_{\rm pe}t \approx 2.63\times 10^4$ ($t\Omega_{\rm ci} \approx 7$) for the M2.0 model (top panels) and the M3.0 model (bottom panels).}
\end{figure*}

The red solid line in Figure \ref{fig:f6} is represented roughly by $\zeta \propto M_{\rm s}^4$ in the range of $2.3 \lesssim M_{\rm s} \leq 3$, but it drops rather abruptly below $2.3$, deviating from the power-law behavior.
We note that the Mach number dependence of $\zeta$ is steeper than that of the ion injection fraction for $Q_\parallel$-shocks, which is roughly $\xi \propto M_{\rm s}^{1.5}$, as shown in Paper I.
This implies that the kinetic processes involved in electron preacceleration might be more sensitive to $M_{\rm s}$ (see Section \ref{sec:s3.2}).

Figure \ref{fig:f6} also shows $\zeta$ for models with $\theta_{\rm Bn} = 53^{\circ}$ (triangles) and $\theta_{\rm Bn} = 73^{\circ}$ (squares).
For shocks with larger $\theta_{\rm Bn}$, the reflection of electrons and the average SDA energy gain are larger, resulting in larger $I$, {\color{black} as shown in Figure \ref{fig:f1} (c) and (d).
Hence, $\zeta$ should be larger at higher obliquity angle.
However, for $\theta_{\rm Bn} > \theta_{\rm limit} \approx 73-78^{\circ}$, $\zeta$ should begin to decrease, as mentioned in Section \ref{sec:s3.2.1}.}

Based on the above results, we propose that the preacceleration of electrons is effective only in $Q_\perp$-shocks with $M_{\rm s} \gtrsim 2.3$ in the hot ICM, that is, $M_{\rm ef}^*\approx 2.3$.
We point out that this is close to the first critical Mach number for ion reflection, $M_{\rm f}^* \approx 2.3$, {\color{black} estimated from the Mach number dependence of the fraction of reflected ions,  $\alpha_{\rm ref,i}$,
shown in Figure \ref{fig:f3}(d)}.
As shown in Figure \ref{fig:f2}, overshoot-undershoot oscillations develop in the shock transition, owing to a sufficient amount of reflected ions, in shocks with $M_{\rm s} \gtrsim 2.3$; with larger magnetic field compression due to the oscillations, more electrons are reflected and energized via SDA (see Section \ref{sec:s3.2.1}).
{\color{black} 
Hence, we expect that the electron reflection is directly linked with the ion reflection, so $M_{\rm ef}^*$ would be related with $M_{\rm f}^*$.}
Note that the critical Mach number, $M_{\rm f}^* \approx 2.3$, is also similar to the first critical Mach number for ion reflection and injection to DSA in $Q_\parallel$-shocks in high-$\beta$ plasmas, $M_{\rm s}^*\approx 2.25$ (Paper I).

\subsection{Upstream Waves}
\label{sec:s4.3}

The nature and origin of upstream waves in collisionless shocks have long been investigated through both analytical and simulation studies with the help of {\it in situ} observations of Earth's bow shock.
In $Q_\parallel$-shocks with low Mach numbers, magnetosonic waves such as phase-standing whistlers and long-wavelength whistlers are known to be excited by backstreaming ions via an ion/ion beam instability \citep[e.g.,][]{kraussvarban1991}.
Especially, in supercritical $Q_\parallel$-shocks, the foreshock region is highly turbulent with large-amplitude waves
and the shock transition can undergo quasi-periodic reformation due to the nonlinear interaction of accumulated waves and the shock front (see Paper I and Figure \ref{fig:f2}(d)).
In $Q_\perp$-shocks, a sufficient amount of incoming protons can be reflected at the shock, which in turn may excite fast magnetosonic waves.
As discussed in Section \ref{sec:s3.1}, two whistler critical Mach numbers, $M_{\rm w}^*$ and $M_{\rm nw}^*$,
are related with the upstream emission of whistler waves and the nonlinear breaking of whistler waves in the shock foot.
{\color{black} In some of our models, $M_{\rm w}^*  < M_{\rm s} < M_{\rm nw}^*$, and hence whistler waves are confined within the shock foot and shock reformation does not occur.}

In {\it supercritical} shocks with $M_{\rm s} \gtrsim 2.3$, we expect to see the following three kinds of waves:
(1) nearly phase-standing whistler waves with $kc/\omega_{\rm pi} \sim 1$ ($kc/\omega_{\rm pe} \sim 0.1$) excited by reflected ions \citep[e.g.,][]{hellinger2007,scholer2007},
where $w_{\rm pi} = \sqrt{4\pi e^{2}n/m_i}$ is the ion plasma frequency ($w_{\rm pi}=0.1 w_{\rm pe}$ for $m_e/m_i = 100$),
(2) phase-standing oblique waves with $kc/\omega_{\rm pe} \sim 0.4$ and larger $\theta_{\rm Bk}$ (the angle between the wave vector $\mathbf{k}$ and $\mathbf{B_0}$) excited by the EFI,
and (3) propagating waves with $kc/\omega_{\rm pe} \sim 0.3$ and smaller $\theta_{\rm Bk}$, also excited by the EFI \citep[e.g.,][]{hellinger2014}.

Here, we focus on the waves excited by the EFI described in Section \ref{sec:s3.2.2}.
Previous studies on the EFI and the EFI-induced waves showed the following characteristics \citep[][GSN14b]{gary2003,camporeale2008,hellinger2014,lazar2014}.
(1) The magnetic field fluctuations in the EFI-induced waves are predominantly along the direction perpendicular to both $\mathbf{k}$ and $\mathbf{B_0}$, i.e., $|\delta B_z|$ is larger than $|\delta B_x|$ and $|\delta B_y|$ in our geometry.
(2) Phase-standing oblique waves with almost zero oscillation frequencies ($\omega_r \approx 0$) have higher growth rates than propagating waves ($\omega_r \ne 0$) .
(3) Nonpropagating modes decay to propagating modes with longer wavelengths and smaller $\theta_{\rm Bk}$.
(4) The EFI-induced waves scatter electrons, resulting in the reduction of electrons temperature anisotropy, 
{\color{black} which in turn leads to the damping of the waves.}

\begin{figure*}[t]
\vskip -0.6cm
\hskip -0.6cm
\centerline{\includegraphics[width=1.2\textwidth]{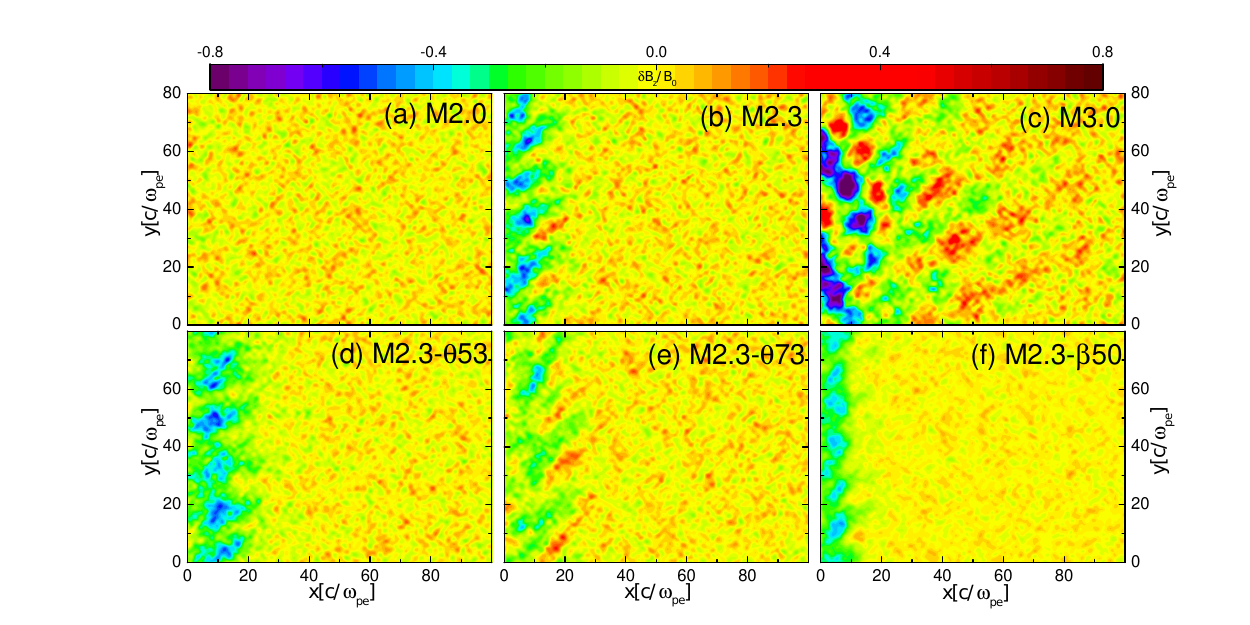}}
\vskip -0.4cm
\caption{\label{fig:f8}
Magnetic field fluctuations, $\delta B_z$, normalized to $B_0$, in the upstream region of $0<(x-x_{\rm sh})w_{\rm pe}/c <100$ at $w_{\rm pe}t \approx 3.76\times 10^4$ ($t\Omega_{\rm ci} \approx 10$) for six different models.}
\end{figure*}

\begin{figure}[t]
\vskip -0.9cm
\hskip 0.1cm
\centerline{\includegraphics[width=0.55\textwidth]{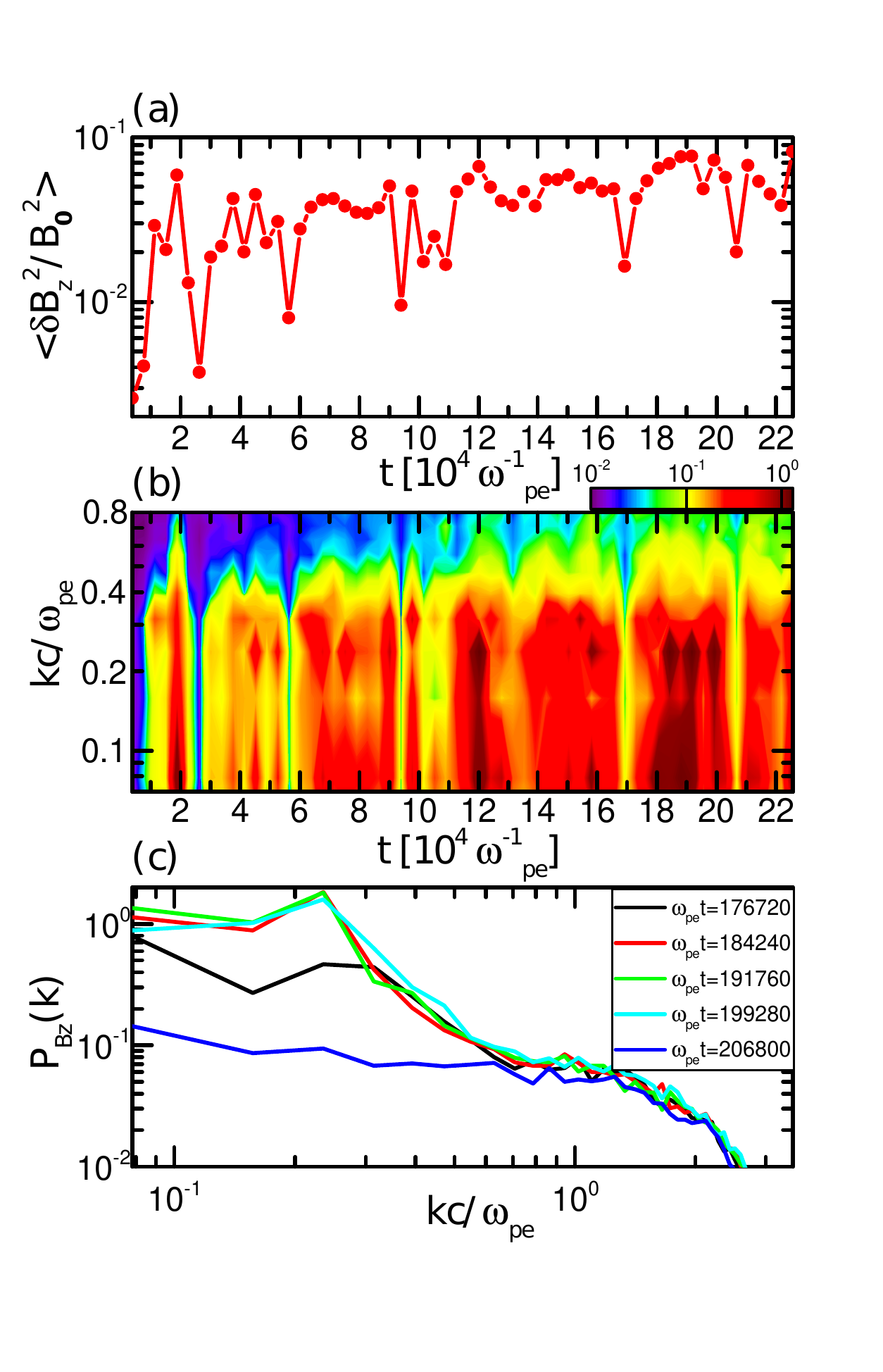}}
\vskip -1.4cm
\caption{\label{fig:f9}
(a) Time evolution of $\left< \delta B_z^2/B_0^2 \right>$, the square of the magnetic field fluctuations normalized to the background magnetic field, averaged over the square region of $80 \times 80 (c/\omega_{\rm pe})^2$ covering $0<(x-x_{\rm sh})w_{\rm pe}/c < 80$, for the M3.0 model.
(b) Time evolution of $P_{B_z}(k) \propto |\delta B_z(k)|^2 k$, the magnetic energy power of $\delta B_z$ in the square region of $80 \times 80 (c/\omega_{\rm pe})^2$ for the M3.0 model.
(c) $P_{B_z}(k)$ versus $kc/w_{\rm pe}$ at five different time epochs.
}
\end{figure}

Figure \ref{fig:f7} shows the distribution of magnetic field fluctuations, $\delta\mathbf{B}$, in the upstream region for the M2.0 and M3.0 models.
The epoch shown, $t\Omega_{\rm ci} \approx 7$ ($ w_{\rm pe} t\approx 2.63\times 10^4$) is early; yet, in the M3.0 model, waves are well developed (see also Figure \ref{fig:f9}), while the energization of electrons is still undergoing (see Figure \ref{fig:f5}). 
{\color{black} For the supercritical shock of the M3.0 model, we interpret that there are ion-induced whistlers in the shock ramp region of $0 \lesssim x-x_s \lesssim 60 c/\omega_{\rm pe}$, while EFI-induced oblique waves are present 
over the whole region shown.
As shown in GSN14b, the EFI-excited waves are oblique with $\theta_{\rm Bk}\sim 60^{\circ}$, and $|\delta B_z| > |\delta B_x|$ and $|\delta B_y|$. 
The increase in $\delta B_y$ toward $x-x_{\rm sh} = 0$ is due to the compression in the shock ramp.
In the subcritical shock of the M2.0 model, on the other hand, the fractions of reflected ions and electrons are not sufficient for either the emission of whistler waves or the excitation of EFI-induced waves, so no substantial waves are present in the shock foot.
This is consistent with the instability condition shown in Figure \ref{fig:f1} (d).}
 
Figure \ref{fig:f8} compares $\delta B_z$ in six different models at $t\Omega_{\rm ci} \approx 10$.
The wave amplitude increases with increasing $M_{\rm s}$, and the EFI seems only marginal in the M2.3 models.
This result confirms our proposal for the ``EFI critical Mach number'' $M_{\rm ef}^*\approx 2.3$, presented in Section \ref{sec:s4.2}.
{\color{black} Moreover, this figure corroborates our findings that the EFI is more efficient at larger $\theta_{\rm Bn}$ and higher $\beta$.}
From $\delta B_z$ of the M2.3 and M3.0 models in Figure \ref{fig:f7} and \ref{fig:f8}, 
the dominant waves in the shock foot seem to have $\lambda \sim 15-20 c/\omega_{\rm pe}$, 
so they are consistent with the EFI-induced waves (GSN14b).

Figure \ref{fig:f9} shows the time evolution of the average of the magnetic field fluctuations, $\left< \delta B_z^2/B_0^2 \right>$, and the magnetic energy power, $P_{Bz}(k) \propto |\delta B_z(k)|^2 k$, of upstream waves for the M3.0 model.
{\color{black} According to the linear analysis by \citet{camporeale2008}, the growth rate of the EFI peaks at 
$k_{\rm max}c/\omega_{\rm pe} \sim 0.4$ for $\beta_{e\parallel}=10$ and $T_{e\perp}/T_{e\parallel}=0.7$.
Thus, we interpret that the powers in the range of $kc/\omega_{\rm pe} \sim 0.2-0.3$ are owing to the oblique waves induced by the EFI, while
those of $kc/\omega_{\rm pe} \lesssim 0.15$ are contributed by the phase-standing whistler waves induced by reflected ions.}

{\color{black} Moreover, through periodic box simulations of the EFI, \citet{camporeale2008} and \citet{hellinger2014} demonstrated that 
initially nonpropagating oblique modes grow and then saturate, followed by the transfer of wave energy into propagating modes with longer wavelengths and smaller $\theta_{\rm Bk}$.
Figure 8 of \citet{camporeale2008} and Figure 4 of \citet{hellinger2014} show that a cycle of the EFI-induced wave growth and decay occurs with the time scales of $t\Omega_{\rm ce} \sim {\rm several}~\times 100$.
We suggest that the oscillatory behaviors of the excited waves with the time scales of $t w_{\rm pe} \sim 2 \times 10^4 - 4 \times 10^4$ 
($t\Omega_{\rm ce} \sim 500-1000$) shown in Figure \ref{fig:f9} would be related to those characteristics of the EFI.
Figure \ref{fig:f9}(c) illustrates such a cycle during the period of $t w_{\rm pe} \approx 1.8-2.1 \times 10^5$: excitation with $k_{\rm max}c/w_{\rm pe} \approx 0.3$ $\rightarrow$ inverse cascade with $k_{\rm max}c/w_{\rm pe} \approx 0.2$ $\rightarrow$ damping of waves.}

{\color{black} Our results indicate that} the EFI-induced waves do not further develop into longer wavelength modes with $\lambda \gg \lambda_{\rm max}$, where $\lambda_{\rm max} \approx 15 - 20 c/\omega_{\rm pe}$ is the wavelength of the maximum linear growth.
Note that $\lambda_{\rm max}$ is close to the gyroradius of electrons with $\gamma \lesssim 2$.
Thus, the acceleration of electrons via {\color{black} resonant scattering by the EFI-induced waves} is saturated.
As a consequence, the energization of electrons stops at the suprathermal stage $(\gamma < 2)$ and does not proceed all the way to the DSA injection momentum $(\gamma_{\rm inj}\approx 10)$.
We interpret that this result should be due to the intrinsic properties of the EFI, rather than the limitations or artifacts of our simulations, 
{\color{black} as shown by the studies of \citet{camporeale2008} and \citet{hellinger2014}.}
Hence, we here conclude that the preacceleration via the EFI alone may not explain the injection of electrons to DSA in weak ICM shocks.
{\color{black} However, the conclusion needs to be further verified through a more detailed study of the EFI and EFI-induced waves for high-$\beta$ ICM plasmas, 
including kinetic linear analyses and numerical simulations, which we leave for a future work.}

\section{Summary}
\label{sec:s5}

In $Q_\perp$-shocks, a substantial fraction of incoming particles are reflected at the shock ramp.
Most of reflected ions are advected downstream along with the underlying magnetic field after about one gyromotin, but yet the structures of the shocks are primarily governed by the dynamics of reflected ions.
Especially in supercritical shocks, the accumulation of reflected ions in the shock ramp generates overshoot-undershoot oscillations in the magnetic field, ion/electron densities, and electric shock potential.
Reflected electrons, on the other hand, can stream along the background magnetic field with small pitch angles in the upstream region. 
As presented in GSN14a and GSN14b, the SDA reflected electrons produce the temperature anisotropy, $T_{e\parallel}> T_{e\perp}$, which induces the EFI; the EFI in turn excites oblique waves in the upstream region.
Electrons are then scattered between the shock ramp and the upstream waves, and gain energies via a Fermi-like process involving multiple cycles of SDA.
All these processes depend most sensitively on $M_{\rm s}$ among a number of shock parameters; for instance, the development of the EFI and the energization of electrons are expected to be inefficient in very weak shocks with $M_{\rm s}$ close to unity.

In this paper, we studied through 2D PIC simulations the preacceleration of electrons facilitated by the EFI in $Q_\perp$-shocks with $M_{\rm s} \lesssim 3$ in the high-$\beta$ ICM.
Various shock parameters are considered, as listed in Table \ref{tab:t1}.
Our findings can be summarized as follows:

1. For ICM $Q_\perp$-shocks, 
{\color{black} ion reflection and overshoot-undershoot oscillations in the shock structures become increasingly more evident for $M_{\rm s} \gtrsim 2.3$,
while the shock structures seem relatively smooth and quasi-stationary for lower Mach number shocks.
Hence we suggest that the effective value of the first critical Mach number would be $M_{\rm f}^*\approx 2.3$,
which is higher than previously estimated from the MHD Rankine-Hugoniot jump condition by \citet{edmiston1984}. 

2. Since electron reflection is affected by ion reflection and the ensuing growth of overshoot-undershoot oscillation,
the EFI critical Mach number, $M_{\rm ef}^*\approx 2.3$, seems to be closely related with $M_{\rm f}^*$.}
Oscillations in the shock structures enhance the magnetic mirror in the shock ramp, providing a favorable condition for the efficient reflection of electrons.
Only in shocks with $M_{\rm s} > M_{\rm ef}^*$, the reflection and SDA of electrons are efficient enough to generate sufficient temperature anisotropies, which can trigger the EFI and the excitation of oblique waves.

3. We presented the fraction of suprathermal electrons, $\zeta(M_{\rm s}, \theta_{\rm Bn})$, defined as the number fraction of electrons with $p \geq p_{\rm spt} = 3.3 p_{\rm th,e}$ in the upstream energy spectrum.
The suprathermal fraction increases with increasing $M_{\rm s}$, roughly as $\zeta \propto M_{\rm s}^4$ for the fiducial models.
Below $M_{\rm ef}^* \approx 2.3$, $\zeta$ drops sharply, indicating inefficient electron preacceleration in low Mach number shocks.
This fraction also increases with increasing $\theta_{\rm Bn}$.
For shocks with larger $\theta_{\rm Bn}$, the reflection of electrons and the average SDA energy gain are larger, and hence $\zeta$ is larger.

4. In the supercritical M3.0 model, the suprthermal tail of electrons extends to higher $\gamma$ in time, but it saturates beyond $t \Omega_{\rm ci} \approx 20$ with the highest energy of $\gamma \lesssim 2$.
In order for suprathermal electrons to be injected to DSA, their energies should reach at least to $\gamma_{\rm inj} \gtrsim 10$.
We interpret that such saturation is due to the lack of wave powers with long wavelengths.
The maximum growth of the EFI in the linear regime is estimated to be at $\lambda_{\rm max}\approx 15 -20 c/\omega_{\rm pe}$.
{\color{black} The EFI becomes stablized owing to the reduction of electron temperature anisotropy,} before waves with $\lambda \gg \lambda_{\rm max}$ develop.
This implies that the preacceleration of electrons due to a Fermi-like process and multiple cycles of SDA, facilitated by the upstream waves excited via the EFI, may not proceed all the way to DSA in high-$\beta$, $Q_\perp$-shocks. 

Our results indicate that processes other than those considered in this paper may be crucial to understand the origin of radio relics in galaxy clusters.
For instance, in the reacceleration model, pre-existing fossil electrons are assumed \citep[e.g.,][]{kang16a,kang16b}.
Especially, fossil electrons with $\gamma \sim 10-100$ could be scattered by ion-induced waves and/or pre-existing turbulent waves and participate to DSA.
\citet{park2015}, for instance, showed through 1D PIC simulations that electrons can be injected to DSA and accelerated via the full Fermi-I process even in $Q_\parallel$ with $M_{\rm A} \approx M_{\rm s} = 20$ and $\theta_{\rm Bn}=30^{\circ}$.
In addition, if shock surfaces are highly non-uniform with varying $M_{\rm s}$ and $\theta_{\rm Bn}$ \citep[e.g.,][]{hong15,ha2018a}, the features of $Q_\perp$ and $Q_\parallel$-shocks may be mixed up, facilitating the upstream environment of abundant waves for electron scattering.
However, all these processes need to be investigated in details before their roles are discussed, and we leave such investigations for future works.

\acknowledgments
{\color{black} The authors thank the anonymous referee for constructive comments.}
H.K. was supported by the Basic Science Research Program of the National Research Foundation of Korea (NRF) through grant 2017R1D1A1A09000567.
D.R. and J.-H. H. were supported by the NRF through grants 2016R1A5A1013277 and 2017R1A2A1A05071429.
J.-H. H. was also supported by the Global PhD Fellowship of the NRF through 2017H1A2A1042370.

\end{document}